\newcommand{\diracslash}[1]{#1\llap{/\kern2pt}}

\newcommand{\be}{\begin{equation}}
\newcommand{\ee}{\end{equation}}
\newcommand{\bea}{\begin{eqnarray}}
\newcommand{\eea}{\end{eqnarray}}
\newcommand{\ba}[1]{\begin{array}{#1}}
\newcommand{\ea}{\end{array}}

\newcommand{\bt}{\begin{tabular}}
\newcommand{\et}{\end{tabular}}
\newcommand{\Tr}{{\rm Tr}}

\newcommand{\pa}{\partial}
\newcommand{\beas}{\begin{eqnarray*}}
\newcommand{\eeas}{\end{eqnarray*}}

\newcommand{\fr}{\frac}

\newcommand{\dg}{\dagger}

\newcommand{\pam}{\partial_\mu}

\documentclass[amsmath,12pt,amssymb,preprint,nofootinbib]{revtex4}
\usepackage{amsfonts} %\usepackage{citesort}
\usepackage{graphicx} % Include figure files
\usepackage{epsfig}
\usepackage{multirow}
\usepackage{bm}% bold math

\begin{document}

\title{Spectral functions of strange vector mesons 
in asymmetric hyperonic matter}

\author{Amruta Mishra}
\email {amruta@physics.iitd.ac.in}
\affiliation
{Department of Physics, Indian Institute of Technology, Delhi,
Hauz Khas, New Delhi -- 110 016, India}
\author {S.P. Misra}
\email {misrasibaprasad@gmail.com}
\affiliation {Institute of Physics, Bhubaneswar -- 751005, India}

\begin{abstract}
We study the medium modifications of the spectral functions 
as well as production cross-sections 
of the strange vector mesons ($\phi$, $K^*$ and $\bar {K^*}$)
in isospin asymmetric strange hadronic matter.
These are obtained from the in-medium masses of the open strange mesons 
and the decay widths $\phi \rightarrow K\bar K$, 
$K^* \rightarrow K\pi$ and $\bar {K^*} \rightarrow {\bar  K}\pi$ 
in the hadronic medium.
The decay widths are computed using a field theoretic model 
of composite hadrons with quark/antiquark constituents,
from the matrix element of the light quark-antiquark pair creation
term of the free Dirac Hamiltonian between the initial and
final states. The matrix element is multiplied with 
a coupling strength parameter for the light quark-antiquark
pair creation, which is fitted to the observed vacuum decay width
of the decay process. There are observed to be substantial modifications 
of the spectral functions as well as production cross-sections 
of these vector mesons due to isospin asymmetry as well as 
strangeness of the hadronic medum at high densities. These studies 
should have observable consequences, e.g. in the yield of the hidden 
and open strange mesons arising from the isospin asymmetric 
high energy heavy ion collisions at the Compressed baryonic matter
(CBM) experiments at the future facility at GSI. 
\end{abstract}

\maketitle

\section{Introduction}
The study of the in-medium properties of the strange mesons is 
relevant for the heavy ion collision experiments, 
as these can affect the experimental observables, e.g.,
the yield, spectra and collective flow of these
mesons in the strongly interacting matter created
from high energy nuclear collisions
%%%%%%%%%%added below Ref.
\cite{C_Hartnack_Phys_Rep_510_2012_119}.
%%%%%%%%%%added above Ref.
%%%%%%%%%%added below 
The understanding of the interactions of the strange mesons with nuclei
and nuclear matter is needed for the study of kaonic atoms
\cite{Friedmann_Gal_Phys_Rep_452_89_2007}, 
the experimental observables of the heavy ion collision experiments,
as well as the composition of matter in the interior of the neutron stars 
\cite{L_Tolos_L_Fabbietti_Prog_Part_Nucl_Phys_112_103770_2020}.
%%%%%%%%%%added above
The possibility of antikaon condensation \cite{kaplan}
in the interior of the neutron stars,
due to the mass drop of the antikaons arising from
attractive interaction with the nucleons, 
initiated a lot of work on the study of in-medium properties 
of these open strange mesons. It is also important to study 
the effects of the isospin asymmetry of the hadronic medium
on the hadron properties in the context
of heavy ion collision experiments as the initial system,
consisting of heavy ions has large asymmetry in the number of
protons and neutrons. The isospin asymmetry effects can affect 
observables like the $\pi^-/\pi^+$ ratio,
the $n/p$ ratio, the $\Delta ^-/\Delta ^{++}$ ratio 
for neutron rich heavy ion collisions.
The effects of the medium modifications of the kaons and antikaons
in the isospin asymmetric hadronic matter can show 
in the experimental observables, 
e.g., the $K^+/K^0$, the $K^-/{\bar {K^0}}$,
the ${K^*}^+/{K^*}^0$ and the ${K^*}^-/{\bar {{K^*}^0}}$
ratios, as well as the yield, spectra and collective flow
of the strange mesons.

The kaons and antikaons in the hadronic medium have 
been studied extensively in the literature.
In Ref. \cite{kaplan}, the interactions of the
kaons and antikaons with the nucleons 
arise due to the leading order vectorial Weinberg-Tomozawa term 
and the attractive kaon-nucleon sigma term at the sub-leading order
within the chiral perturbation theory. 
The $K$ and $\bar K$ mesons have also been studied using a
meson exchange model
\cite{Glendenning_Schaffner_PRC_60_025803_1999,Debades_PRC86_045803_2012}
incorporating the interactions of these open 
strange mesons to the meson fields ($\sigma$, $\omega$, $\rho$)
similar to the interactions of the baryons to these meson fields
in Quantum Hadrodynamics (QHD) model
\cite{Serot_Rep_Prog_Phys_55_1855_1992}.
Within the framework of the Quark meson coupling (QMC) model, 
the light ($q=(u,d)$) quark (antiquark) constituents of a hadron, 
$h$, confined inside a bag of radius $R_h(R_h^*)$ in the vacuum 
(hadronic medium), interact via the exchange of the light ($\sigma$, 
$\omega$, $\rho$) mesons. The scalar potential, $V_s^h$, 
experienced by the hadron in the medium, 
which is equal to the mass shift ($m_h^*-m_h$) of the hadron, 
has dominant contributions from the scalar potentials 
($V_\sigma^q=g_\sigma^q\sigma$) experienced by the light 
($u$, $d$) quark (antiquark) constituents in the medium. 
There are additional negligible contributions to the mass 
shift of the hadron due to the center of mass and gluonic 
fluctuation effects taken into account by a parameter, $z_h$
in the QMC model \cite{qmc_phi,Krein_Prog_Part_Nucl_Phys}.
The mass shifts of the hadrons with the same light ($q=u,d$) 
quark (antiquark) constituents e.g., for $K(\bar K)$ 
and $K^*(\bar {K^*})$ mesons, are thus observed 
to be very close to each other, in spite of
the fact that $K$ and $\bar K$ are pesudoscalar mesons
with spin zero, whereas $K^*$ and $\bar {K^*}$ mesons have
spin, $S$=1. It might, however, be mentioned here that
in order to reproduce the empirically extracted replusive
$K^+ N$ total potential from the repulsive $K^+$-Nucleus interaction 
\cite{qmc_phi}, the vector potential $V_\omega^q$ of the
constituent light quark (antiquark) of $K(\bar K$) has to be 
increased by a factor of $(1.4)^2$ in the QMC model 
\cite{qmc_phi}.
Within the framework of the coupled channel approach, 
the resonance $\Lambda(1405)$
is generated dynamically from the $\bar K N$ interaction
using the lowest order Weinberg-Tomozawa interaction
in a chiral Lagrangian \cite{Oset_RamosNPA635_1998_99}.
The $\bar K N$ scattering amplitude in the hadronic medium
is obtained by the self-consistent solution of the coupled channel 
Lippmann Schwinger equations, including the medium effects from
the Pauli blocking, the mean field binding potentials for the baryons 
and the sel-energies of $\bar K$ and $\pi$
\cite{A_Ramos_E_Oset_NPA_671_481_2000}.
For the $K N$ interaction, due to absence of any resonance
near to the threshold, the $T\rho$ approximation is a good approximation 
for the study the self-energies of the kaons 
at low densities, and, a self-consistent calculation of 
the scattering amplitude has only small modifications 
to the results obtained using this approximation.
Using the Parton-Hadron-String-Dynamics (PHSD) transport model,
and, the in-medium spectral functions of the pseudoscalar 
($K$, $\bar K$) mesons, as well as, of the vector ($K^*$ and 
$\bar {K^*}$) mesons, calculated using the self-consistent 
coupled channel approach, the dynamics of $K^*$ and $\bar {K^*}$
in heavy ion collision experiments have been studied
in Refs. \cite{Elena_16,Elena_17}. 
These calculations show that the production of
the strange vector $K^*$($\bar {K^*}$) mesons 
by hadronization from the QGP phase is quite small 
at RHIC as well as at LHC and these vector mesons are dominantly
produced at the later hadronic stage from $K(\bar K)\pi$ 
scatterings. In Ref. \cite{Elena_17}, the PHSD calculations
have also been performed for heavy ion collision experiments 
at low beam energies relevant for the FAIR and NICA conditions, 
where the medium effects may become visible due to the
large baryon density matter which will be created at these
facilities. 

In the present work, we study the relativstic Breit-Wigner 
spectral functions as well as production cross-sections 
for the vector mesons, $K^*$, $\bar {K^*}$ and 
$\phi$ mesons in isospin asymmetric strange hadronic
matter. These are obtained from the mass modifications 
of the open strange mesons and the in-medium decay widths
%%added below
 for the processes
%%added above
$\phi \rightarrow K\bar K$, $K^*\rightarrow K\pi$
and ${\bar {K^*}}\rightarrow {\bar K}\pi$ calculated 
from the mass modifications
of the $K(\bar K)$ and $K^* (\bar {K^*})$ mesons. 
%These studies can have implications on the experimental
%observables, e.g, the production of the hidden ($\phi$) 
%and open strange ($K (\bar K)$ and $K^*(\bar {K^*}$)) mesons.
The masses of the pseudocalar open strange mesons
(kaons and antikaons) in the hadronic medium 
\cite{kmeson1,isoamss,isoamss1,isoamss2}, have been
calculated using a chiral SU(3) model \cite{paper3,kristof1}. 
The mass modifications of the $K$ and $\bar K$ mesons 
in the isospin asymmetric hyperonic matter 
result from their interactions with the
baryons and the scalar mesons. 
%%%%%%%%modified below
In the chiral SU(3) model, the interactions of the kaons 
and antikaons are due to the vectorial interaction with the baryons
given by the Weinberg--Tomozawa term at the leading order,
as well as, the scalar exchange and the range terms 
at the next to leading order in the chiral perturbation
theory.
%%%%%%%%%%%%%%%%%%%%%%%%%%%%%%%%%%%%%%%%%%%%%%%%%
%%%%%modified the sentences below
%%%%%%%%%%%%%%%%%%%%%%%%%%%%%%%%%%%%%%%%%%%%%%%%%
The $K^* (\bar {K^*})$ and the $K(\bar K)$ mesons
are observed to have very similar mass modifications 
in the medium in the QMC model, 
due to the identical light quark (and antiquark) 
constituents, i.e., $q\bar s (s \bar q),\; q=(u,d)$, 
of these mesons \cite{qmc_phi,Krein_Prog_Part_Nucl_Phys}.
In the present work, we assume the mass shifts of the
vector $K^*$ and $\bar {K^*}$ mesons to be the same as the 
shifts in the masses of the pseudoscalar mesons $K$ 
and $\bar K$ mesons, the latter being calculated 
using the chiral SU(3) model \cite{isoamss2}. 
%%%%%%%%%%%%%%%%%%%%%%%%%%%%%%%%%%%%%%%%%%%%%%%%%
%%%%%modified the sentences  above
%%%%%%%%%%%%%%%%%%%%%%%%%%%%%%%%%%%%%%%%%%%%%%%%%
The spectral functions as well
as the production cross-sections of the vector mesons 
$\phi$, $K^*$ and $\bar {K^*}$ in the strange hadronic medium 
computed in the present work
are observed to have significant modifications due to isospin asymmetry
as well as strangeness fraction at high densities. These should 
have observable consequences on the production of the hidden 
and open strange mesons
%%%%%%%%%%added Ref. below
\cite{C_Hartnack_Phys_Rep_510_2012_119}
%%%%%%%%%%added Ref. above
in asymmetric heavy ion collisions at the compressed
baryonic matter (CBM) experiments planned at the future facility
at GSI.

The outline of the paper is as follows: In subsection IIA, we 
discuss briefly the study of the in-medium masses of the
kaons and antikaons in isospin asymmetric strange hadronic matter
within a chiral SU(3) model. 
The in-medium decay widths of the $\phi$ meson to $K\bar K$ 
and of the $K^* (\bar {K^*})$ meson to $K (\bar K)\pi$, 
are calculated from the mass modifications 
of the open strange mesons. These decay widths
are calculated using a field theoretical model of composite 
hadrons with quark (and antiquark) constituents as described 
in  subsection IIB. 
Subsection IIC gives a description of the relativistic Breit-Wigner 
spectral functions as well as the production cross-sections of the 
vector mesons, $K^*$, $\bar {K^*}$ and $\phi$ as studied 
in the present work.
In section III, we discuss the effects of isospin asymmetry,
strangeness and density on the spectral properties
and the production cross-sections of the strange vector mesons.
Section IV summarizes the findings of the present investigation. 

\section{Strange mesons in asymmetric hyperonic matter}
\subsection{\bf {MASSES}:}

The isospin asymmetric strange hadronic matter is described
by a chiral SU(3) model based on a nonlinear realization 
of the chiral symmetry \cite{weinberg,coleman,bardeen} 
and the broken scale invariance \cite{sche1,ellis}.
The model describes the interactions of the baryons and the mesons
(scalar, vector, pseudoscalar and axialvector) and the
dilaton field, $\chi$, which is incorporated in the model
to mimick the scale symmetry breaking of Quantum 
Chromodynamics (QCD). The model has been described in detail
in Refs. \cite{paper3,kristof1}.
%%%%%%added below in response to comment 2 of 1st referee 
The general form of the Lagrangian density is given as 
\cite{paper3}
\bea
{\cal L}  = {\cal L}_{kin} + \sum_{W}{\cal L}_{BW}
          +  {\cal L}_{vec} + {\cal L}_0 +
{\cal L}_{scalebreak}+ {\cal L}_{SB},
\label{genlag} \eea
where, $ {\cal L}_{kin} $ corresponds to the kinetic energy terms
of the baryons and the mesons,
${\cal L}_{BW}$ contains the baryon-meson interactions,
$ {\cal L}_{vec} $ describes the dynamical mass
generation of the vector mesons via couplings to the scalar fields
and contains additionally quartic self-interactions of the vector
fields, ${\cal L}_0 $ contains the meson-meson interaction terms
%inducing the spontaneous breaking of chiral symmetry, 
${\cal L}_{scalebreak}$ is a scale invariance breaking logarithmic
potential, $ {\cal L}_{SB} $ describes the explicit chiral symmetry
breaking.
%%%%%%%%added below to show explicitly the various terms
The kinetic energy terms are given as
\bea
\label{kinetic}
{\cal L}_{kin} &=& i \Tr \overline{B} \gamma_{\mu} D^{\mu}B
                + \frac{1}{2} \Tr D_{\mu} X D^{\mu} X
+  \Tr (u_{\mu} X u^{\mu}X +X u_{\mu} u^{\mu} X)
                + \frac{1}{2}\Tr D_{\mu} Y D^{\mu} Y \nonumber \\
               &+&\frac {1}{2} D_{\mu} \chi D^{\mu} \chi
                - \frac{ 1 }{ 4 } \Tr
\left( V_{ \mu \nu }  V^{\mu \nu }  \right)
- \frac{ 1 }{ 4 } \Tr \left( {\cal A}_{ \mu \nu } {\cal A}^{\mu \nu }
 \right)
- \frac{ 1 }{ 4 } \Tr \left(F_{ \mu \nu } F^{\mu \nu }  \right),
\eea
where, $B$ is the baryon octet, $X$ is the scalar meson
multiplet, $Y$ is the pseudoscalar chiral singlet, 
$\chi$ is the scalar dilaton field,
${V}_{\mu\nu}=\pa_\mu{V}_\nu-\pa_\nu{V}_\mu$,
${\cal A}_{\mu\nu}= \pa_\mu{\cal A}_\nu-\pa_\nu{\cal A}_\mu $, 
and $F_{\mu\nu}$ are the field strength tensors of 
the vector meson multiplet, $V^\mu$, 
the axial vector meson multiplet ${\cal A}^\mu$ 
and the photon field, $A^\mu$.
In equation (\ref{kinetic}), 
$u_\mu= -\fr{i}{4}
[(u^\dg(\pam u)-(\pam u^\dg) u)
 - (u (\pam u^\dg)-(\pam u) u^\dg)]$,
where, $u=\exp\Bigg[\fr{i}{\sigma_0}\pi^a\lambda^a\gamma_5\Bigg]$.
The covariant derivative of a field $\Phi (\equiv B,X,Y,\chi)$
reads $ D_\mu \Phi = \pam\Phi + [\Gamma_\mu,\Phi]$, with
$\Gamma_\mu=-\fr{i}{4} [(u^\dg(\pam u)-(\pam u^\dg) u)
 + (u (\pam u^\dg)-(\pam u) u^\dg)]$.

${\cal L}_{BW}$ in equation (\ref{genlag}) describes
the interactions of the baryons with the meson, $W$ 
%(scalar, pesudoscalar, vector, axialvector meson) 
and has the general form \cite{paper3} 
\begin{eqnarray}
{\cal {L}_{BW}} = {\sqrt 2 } g_8^W \Big ( \alpha_W [\bar B O B W]_F
+(1-\alpha_W) [\bar B O B W]_D\Big) 
- g_1 ^W \frac{1}{\sqrt 3} {\rm {Tr} }(\bar B O B) {\rm Tr} W,
\label{L_BW}
\end{eqnarray}
with $[\bar B O B W]_F={\rm {Tr}}(\bar B O W B-\bar B O B W)$ and
$[\bar B O B W]_D={\rm {Tr}}(\bar B O W B+\bar B O B W)
-\frac{2}{3} Tr (\bar B O B) {\rm {Tr}} W$. 
The above Lagrangian describes the interactions of the baryons 
with scalar mesons ($W=X$, $O$=1), with vector mesons 
($W=V_\mu$, $O=\gamma^\mu$),
with axial vector mesons ($W={\cal A}_\mu$, $O=\gamma^\mu \gamma^5$) 
and with pseudoscalar mesons ($W=u_\mu$, $O=\gamma^\mu \gamma^5$).
%%%%%added above in response to comment 2 of 1st referee 
The hadronic matter is described using the mean field
approximation, which replaces the meson fields with their
expectation values. 
%%%%%added below in response to comment 2 of 1st referee 
Within this approximation, every scalar field, 
$\phi (x)  \rightarrow \langle \phi \rangle \equiv \phi$ and 
vector field, $V^\mu (x) \rightarrow \delta ^{\mu 0} 
\langle V^\mu\rangle \equiv \delta^{\mu 0} V^0$.
Assuming the hadronic matter to be uniform, static
as well as rotationally invariant system, the expectation values 
are independent of space-time and $\langle V^i\rangle=0$
\cite{Serot_Rep_Prog_Phys_55_1855_1992}.
The expectation values of the isovector scalar
and isovector vector meson fields, $\delta^a$ and 
$\rho^{\mu a}$ with $a=1,2,3$, are replaced by
$\delta^a \delta ^{a3}$ and the $\rho^{\mu a}\delta ^{\mu 0}\delta^{a3}$,
as there is contribution only from the neutral scalar and vector mesons
for the ground state. 
The baryon-scalar meson interaction generate the masses
of the baryons and the experimentally observed vacuum masses are fitted
from the parameters $\alpha_s$, $g_1^S$ and $g_8^S$.
In the mean field approximation, the baryon-meson interaction terms
have contributions only from the scalar mesons and the vector 
mesons and are given as
\begin{equation}
{\cal L}_{BS}+{\cal L}_{BV}=
-\sum_i \bar {\psi_i}[m_i^* +g_{\omega i}\gamma^0 \omega 
+g_{\rho i}\gamma^0 \rho +g_{\phi i}\gamma^0 \phi]\psi_i.
\end{equation}
The baryon-scalar meson interaction term, 
${\cal L}_{BS}$ generates the masses of the baryons 
in the octet. The effective mass of the $i$-th baryon
($i=p,n,\Sigma^\pm,\Sigma^0,\Xi^0,\Xi^-$),
described by the fermion field $\psi_i$, 
is given as
\begin{equation}
m_i^*=-(g_{\sigma i}\sigma +g_{\zeta i}\zeta +g_{\delta i}\delta),
\label{meffbaryon}
\end{equation}
where $\sigma$, $\zeta$ and $\delta$ are the expectation values
of the non-strange scalar isoscalar field, $\sigma$,
strange scalar isoscalar field $\zeta$ and third component
of the non-strange scalar isovector $\delta$ fields respectively.
%The mass modifications of the baryons thus arise due to
%the scalar potentials.
The fields $\sigma$, $\zeta$ and $\delta$ may be identified
as the particles $f_0(500)$, $f_0(980)$ and $a_0(980)$
of the Particle Data Group \cite{pdg_2020}.
%%%%%added above in response to comment 2 of 1st referee 
The expectation values of the meson fields are determined 
by solving the coupled equations of motion of these fields 
\cite{kristof1} in the mean field approximation,
additionally assuming that 
\begin{eqnarray}
 \bar \psi_i \psi_j \rightarrow \delta_{ij} \langle \bar \psi_i \psi_i
\rangle \equiv \delta_{ij} \rho_i^s, \;\;\;
 \bar \psi_i \gamma^\mu \psi_j \rightarrow \delta_{ij} \delta^{\mu 0}
\langle \bar \psi_i \gamma^ 0 \psi_i
\rangle \equiv \delta_{ij} \delta^{\mu 0} \rho_i,
\label{vec_scalar_densities}
\end{eqnarray}
where, $\rho_i^s$ and $\rho_i$ are the scalar and number 
densities of baryon of species $i$.
%%%%deleted the part below as it is mentioned already above
%%%as part of response to comment 2 of Ref. 1
% described by the fermion field $\psi_i$, 
%%with $i=p,n,\Lambda,\Sigma^+,\Sigma^0,\Sigma^-$,$\Xi^-,\Xi^0$.
%%%%deleted the part above as it is  already mentioned
%%%as part of response to comment 2 of Ref. 1
These are calculated for given values of the isospin asymmetry parameter,
$\eta= -\frac{\sum_{i}  {I_{3i} \rho_{i}}}{\rho_{B}}$,
and the strangeness fraction, $f_s=
\frac{\sum_{i} {s_{i} \rho_{i}}}{\rho_{B}}$,
where $I_{3i}$ is the third component of the isospin of the
baryon of species, $i$ and  $s_{i}$ is the number 
of strange quarks in the $i$-th baryon.
%%%%added below in reponse to 1st comment of the 2nd referee
In the isospin asymmetric hadronic medium, the expectation
value of the third component of the scalar-isoscalar field 
$\delta (\sim (\bar u u -\bar d d))$ becomes non-zero,
and leads to the masses of the baryons (given by equation 
(\ref{meffbaryon})) within a given isospin multiplet, 
to be different, determined by $g_{\delta i}$, 
the coupling of the $i$-th baryon with the 
field, $\delta$. 

%%%%added above in reponse to 1st comment of the 2nd referee

The modifications of the masses of the kaon and antikaon
in the isospin asymmetric strange hadronic matter 
are calculated from the interactions of these mesons
with the baryons and the scalar mesons in the chiral SU(3) model.
These interactions include the Weinberg-Tomozawa term at the
leading order, as well as the scalar exchange and the range terms
in the next to leading order of chiral perturbation theory.
The dispersion relations of the $K$ and $\bar K$ mesons
are obtained by Fourier transformations of the equations of
motion of these mesons \cite{isoamss2}. These are given as
\begin{equation}
-\omega^2+ {|{\bf k}|}^2 + m_K^2 -\Pi_{K(\bar K)}(\omega, |{\bf k}|)=0,
\label{disp}
\end{equation}
where $\Pi (\omega, |{\bf k}|)$ denotes 
the kaon (antikaon) self energy in the medium.
Explicitly, the self energy $\Pi (\omega,|{\bf k}|)$ for the kaon doublet,
($K^+$,$K^0$) is given as
\begin{eqnarray}
\Pi_{K} (\omega, |{\bf k}|) &= & -\frac {1}{4 f_K^2}\Big [3 (\rho_p +\rho_n)
\pm (\rho_p -\rho_n) \pm 2 (\rho_{\Sigma^+}-\rho_{\Sigma^-})
\nonumber \\
&-&\big ( 3 (\rho_{\Xi^-} +\rho_{\Xi^0}) \pm (\rho_{\Xi^-} -\rho_{\Xi^0})
\big)
\Big ] \omega\nonumber \\
&+&\frac {m_K^2}{2 f_K} (\sigma ' +\sqrt 2 \zeta ' \pm \delta ')
\nonumber \\ & +& \Big [- \frac {1}{f_K}
(\sigma ' +\sqrt 2 \zeta ' \pm \delta ')
+\frac {d_1}{2 f_K ^2} (\rho^s_p +\rho^s_n
\nonumber \\
&+&\rho^s_{\Lambda^0}+\rho^s_{\Sigma^+}+\rho^s_{\Sigma^0}
+\rho^s_{\Sigma^-}+\rho^s_{\Xi^-}+\rho^s_{\Xi^0}
)\nonumber \\
&+&\frac {d_2}{4 f_K ^2} \Big ((\rho^s_p+\rho^s_n)
\pm (\rho^s_p -\rho^s _n)
+\rho^s_{\Sigma ^0}+\frac {5}{3} \rho^s_{\Lambda^0}
\nonumber \\
&+& (\rho^s_{\Sigma ^+}+\rho^s_{\Sigma ^-})
\pm (\rho^s_{\Sigma ^+}-\rho^s_{\Sigma ^-})
+  2 \rho^s_ {\Xi^-}+
2 \rho^s _ {\Xi^0}
\Big )
\Big ]
(\omega ^2 - |{\bf k}|^2),
\label{selfk}
\end{eqnarray}
where the $\pm$ signs refer to the $K^+$ and $K^0$ respectively.
In the above, $\sigma'(=\sigma-\sigma _0)$,
$\zeta'(=\zeta-\zeta_0)$ and  $\delta'(=\delta-\delta_0)$
are the fluctuations of the scalar-isoscalar fields $\sigma$ and $\zeta$,
and the third component of the scalar-isovector field, $\delta$,
from their vacuum expectation values.
The vacuum expectation value of $\delta$ is zero ($\delta_0$=0), since
a nonzero value for it will break the isospin symmetry of the vacuum.
The fluctuation of the dilaton field 
in the medium has been neglected in the present work,
as it is related to the gluon condensate, whose modification
in the hadronic medium is negligible.

Similarly, for the antikaon doublet, ($K^-$,$\bar {K^0}$),
the self energy is calculated as \cite{isoamss2}
\begin{eqnarray}
\Pi_{\bar K} (\omega, |{\bf k}|) &= & \frac {1}{4 f_K^2}\Big [3 (\rho_p +\rho_n)
\pm (\rho_p -\rho_n) \pm 2 (\rho_{\Sigma^+}-\rho_{\Sigma^-})
\nonumber \\
&-& \big ( 3 (\rho_{\Xi^-} +\rho_{\Xi^0}) \pm (\rho_{\Xi^-} -\rho_{\Xi^0})
\big)
\Big ] \omega\nonumber \\
&+&\frac {m_K^2}{2 f_K} (\sigma ' +\sqrt 2 \zeta ' \pm \delta ')
\nonumber \\ & +& \Big [- \frac {1}{f_K}
(\sigma ' +\sqrt 2 \zeta ' \pm \delta ')
+\frac {d_1}{2 f_K ^2} (\rho^s_p +\rho^s_n
\nonumber \\
&+&\rho^s_{\Lambda^0}+\rho^s_{\Sigma^+}+\rho^s_{\Sigma^0}
+\rho^s_{\Sigma^-}+\rho^s_{\Xi^-}+\rho^s_{\Xi^0}
)\nonumber \\
&+&\frac {d_2}{4 f_K ^2} \Big ((\rho^s_p +\rho^s_n)
\pm   (\rho^s_p -\rho^s_n)
+\rho^s_{\Sigma ^0}+\frac {5}{3} \rho^s_{\Lambda^0}
\nonumber \\
&+& (\rho^s_{\Sigma ^+}+\rho^s_{\Sigma ^-})
\pm (\rho^s_{\Sigma ^+}-\rho^s_{\Sigma ^-})
 +  2 \rho^s_{\Xi^-}+ 2 \rho^s_{\Xi^0}
\Big )
\Big ]
(\omega ^2 - |{\bf k}|^2),
\label{selfkbar}
\end{eqnarray}
where the $\pm$ signs refer to the $K^-$ and $\bar {K^0}$ respectively.
In the equations (\ref{selfk}) and (\ref{selfkbar}), the first term 
is due to the vectorial Weinberg-Tomozawa term, the second term 
is the contribution due to the scalar exchange, and, the last 
three terms are due to the range terms.
The parameters $d_1$ and $d_2$ of the range terms are determined 
to be 2.56/$m_K$ and 0.73/$m_K$, by a fit of the empirical values 
of the KN scattering lengths for I=0 and I=1 channels 
\cite{isoamss,isoamss1,isoamss2}, taken as
$a _{KN} (I=0) \approx -0.09 ~ {\rm {fm}}$ and 
$a _{KN} (I=1) \approx -0.31 ~ {\rm {fm}}$
\cite{thorsson,juergen,barnes}.
%%%%%%added below in response to the 1st comment of the 2nd referee
The masses of the kaons and antikaons in the medium,
$m_K^*$ and $m_{\bar K}^*$ defined as the energies 
of these mesons for $|{\bf k}|=0$, 
%i.e., $m_{K(\bar K})^*=\omega (|{\bf k}|=0)$.   
are obtained from the solution of the dispersion relation 
(\ref{disp}) for $|{\bf k}|=0$ 
%%%%%%%%added new Ref.
\cite{C_Hartnack_Phys_Rep_510_2012_119}.
%%%%%%%%added new Ref.
In the early work based on chiral 
perturbation theory \cite{kaplan}, the masses of the kaons
and antikaons in the nuclear medium were obtained 
by solving the dispersion relation for $|{\bf k}|=0$,
using the leading order vectorial Weinberg-Tomozawa term
and the attractive $KN$ sigma term at the sub-leading order, 
the latter largely responsible for the possibility of  
anikaon condensation in the interior of neutron stars
due to drop of the mass of the antikaons
in nuclear matter. The Weinberg-Tomozawa term leads 
to an attraction (repulsion) for the  kaons (antikaons)
in nuclear matter. 
In the chiral effective model used in the present work, 
the mass modifications of the kaons and antikaons
arise from their interactions with the baryons 
and the scalar mesons retaining the leading order
vectorial Weinberg-Tomozawa interaction, as well as,
the scalar exchange and the range terms
at the next to leading order. For the isospin symmetric
hadronic matter ($\eta$=0), as has already been mentioned,
$\delta$=0, the scalar fields $\sigma$ and $\zeta$ 
are obtained from their equations of motion.
For the isospin asymmetric hadronic medium,
($\eta \ne 0$), the expectation values of the scalar fields 
are obtained by solving the coupled equations for
$\sigma$, $\zeta$ and $\delta$, which are used in the
dispersion relation given by (\ref{disp}) to obtain 
the in-medium masses of the $K$ and $\bar K$ mesons 
for given values of the baryon density, $\rho_B$, the 
isospin asymmetry parameter, $\eta$ and the strangeness 
fraction, $f_s$. 
%%%%%%added above in response to the 1st comment of the 2nd referee

The in-medium decay widths of $\phi \rightarrow K\bar K$
and $K^* (\bar {K^*}) \rightarrow K (\bar K)\pi$ are calculated from 
the mass modifications of the open strange mesons.
The in-medium masses of the kaons and the antikaons
have been calculated in isospin asymmetric (nuclear) hyperonic matter,
using the chiral SU(3) model in Ref. \cite{isoamss2}.
%%%%%%%%%%%%added below in response to comment 3 of Ref. 2
It might be noted here that the interactions of the open 
strange vector mesons, $K^*$ and $\bar {K^*}$ with the baryons, 
using the Lagrangian ${\cal L}_{BW}$,
for $W=V$, turn out to be of the form 
$\bar \psi^i \gamma^\mu \psi^j V_\mu$, with $i\ne j$,
which vanish due to the assumption 
$\bar \psi_i \gamma^\mu \psi_j \rightarrow \delta_{ij} \delta^{\mu 0}
\langle \bar \psi_i \gamma^ 0 \psi_i
\rangle \equiv \delta_{ij} \delta^{\mu 0} \rho_i$,
as given by equation (\ref{vec_scalar_densities}). 
The only vector meson fields which have non-zero
contributions from the baryon-meson interaction terms
given by equation (\ref{L_BW}) are the
$\omega$, $\rho_0$ and $\phi$ mesons. 
The calculation of the mass modifications 
of the $K^*$ and $\bar {K^*}$ mesons using the chiral effective model
is thus beyond the scope of the present work. 
%%%%%%%%%%%%added above in response to comment 3 of Ref. 2
As has already been mentioned, in the present work, 
we assume the mass modification 
of the vector open strange meson, $K^*$ ($\bar {K^*}$) 
to be same as for the $K(\bar K)$ meson,
i.e, $\Delta m_{K^*(\bar {K^*})} \equiv m_{K^*(\bar {K^*})}^{*}
-{m_{K^* (\bar {K^*})}^{vac}}
={m_{K(\bar K)}^{*}}-{m_{K (\bar K)}^{vac}}$.
The decay widths of $\phi \rightarrow K\bar K$,
$K^* (\bar {K^*}) \rightarrow K (\bar K)\pi$ 
are calculated using a field theoretical model of hadrons 
with quark (and antiquark) constituents as described in the 
following subsection. 
\begin{figure}
\vspace{-1.4cm}
\begin{center}
%\begin{tabular}{c c }
%\includegraphics[width=16cm,height=16cm]{mkkbar_dens.eps}
\includegraphics[width=18cm,height=20cm]{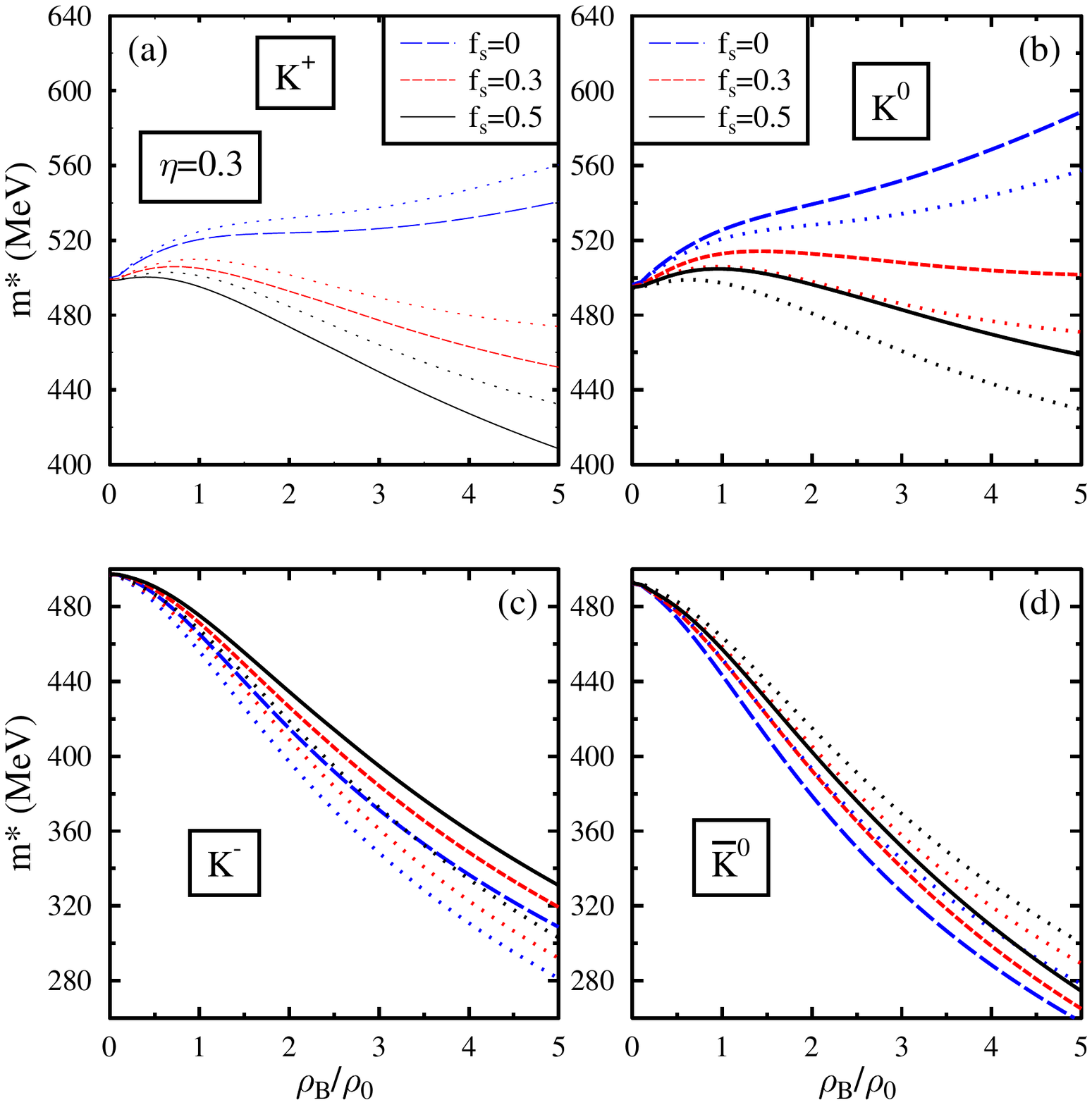}
%\centering
%\psfig{file=fig1.eps,width=9cm}
\caption{
The masses of $K(K^+,K^0)$ and $\bar K(K^-,\bar {K^0})$ mesons
in the (strange) hadronic matter
with strangenss fractions, $f_s=0,0.3,0.5$ for isospin asymmetric 
parameter $\eta$=0.3. These are compared with the isospin symmetric
case ($\eta$=0) shown as dotted lines. 
\label{mkkbar_dens}
}
\end{center}
\end{figure}

%\end{document}
\begin{figure}
%\vspace{-1.4cm}
%\hspace{-0.8cm}
\begin{center}
%\begin{tabular}{c c }
%\includegraphics[width=16cm,height=16cm]{dwFT_Kstr_dens_rev1.eps}
%\includegraphics[width=11.5cm,height=11.5cm]{fig2.pdf}
\includegraphics[width=20cm,height=20cm]{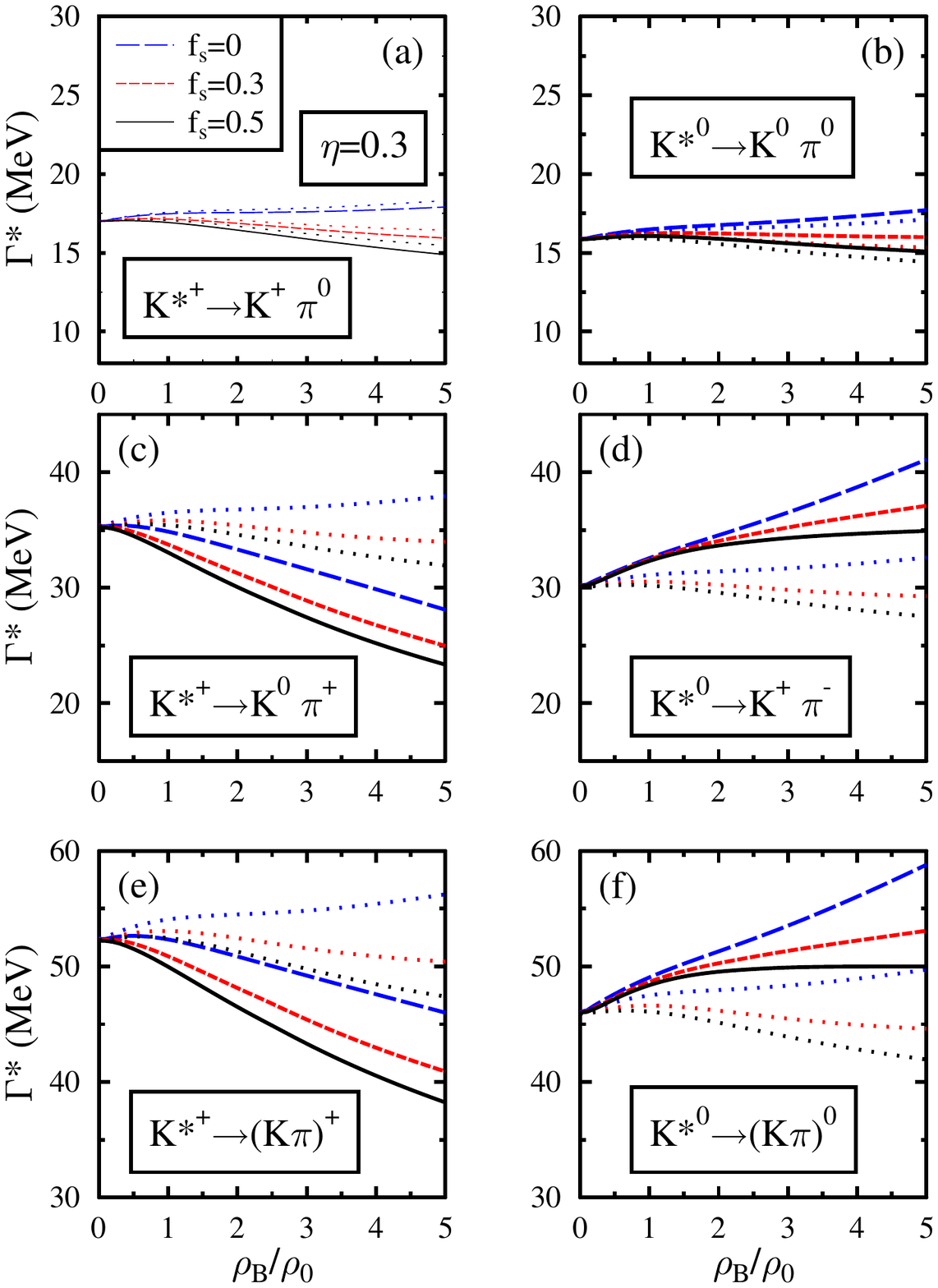}
\caption{
Decay widths of the charged open strange vector meson, 
${K^*}({K^*}^+,{K^*}^0)$ to $(K\pi)^{+,0}$ for $f_s=0,0.3,0.5$ 
for isospin asymmetric matter with $\eta$=0.3. 
These are compared 
to the results obtained for symmetric matter ($\eta$=0)
shown as dotted lines.  
\label{dwFT_Kstr_dens_rev1}
}
\end{center}
\end{figure}
\begin{figure}
\vspace{-1.4cm}
\hspace{-0.8cm}
\begin{center}
%\begin{tabular}{c c }
%\includegraphics[width=16cm,height=16cm]{dwFT_Kstrbar_dens_rev1.eps}
\includegraphics[width=20cm,height=20cm]{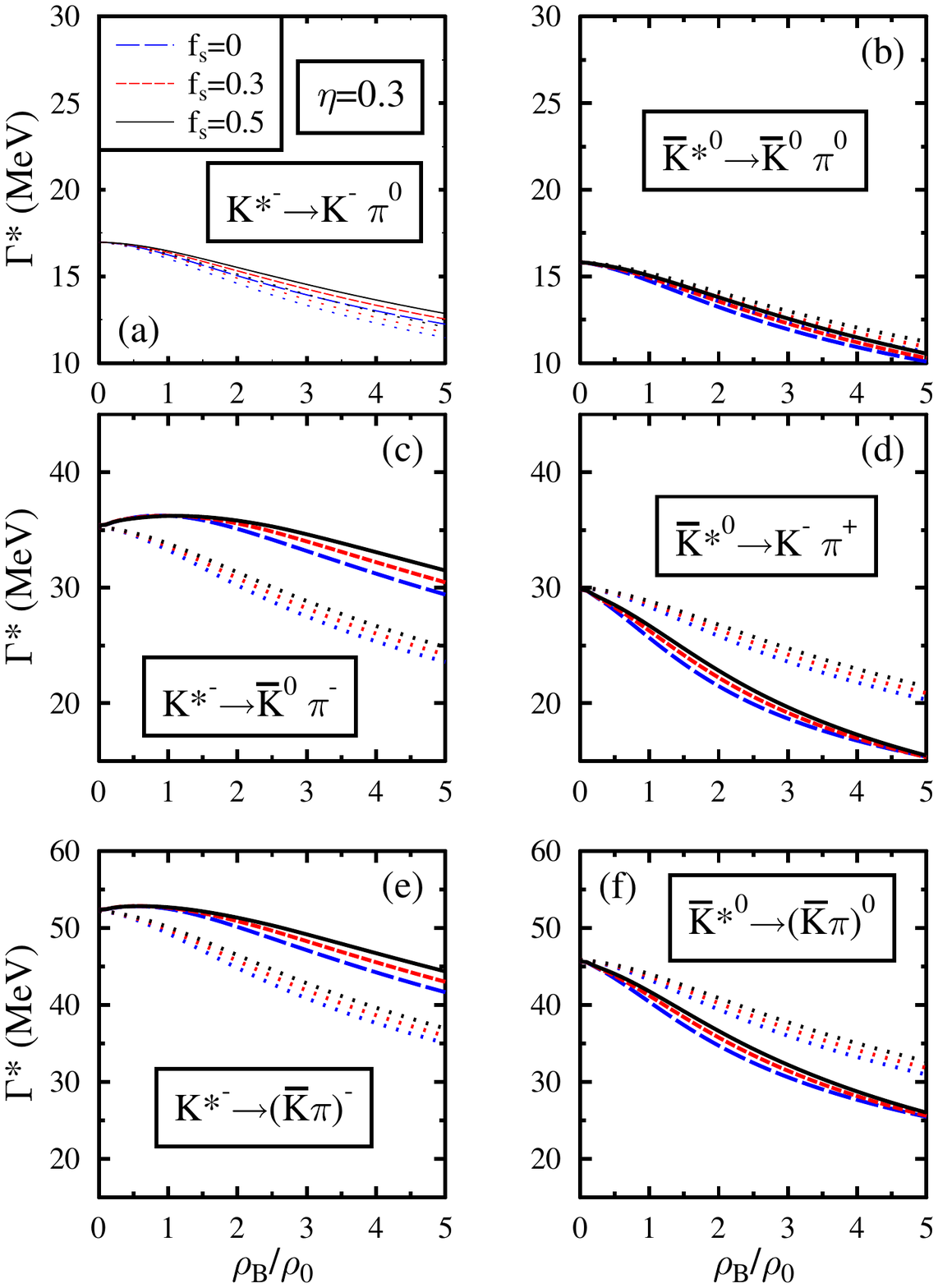}
\caption{
Decay widths of the charged open strange vector meson, 
$\bar {K^*}({K^*}^-,\bar {{K^*}^0})$
to $({\bar K}\pi)^{-,0}$ for $f_s=0,0.3,0.5$ 
for isospin asymmetric matter with $\eta$=0.3.
These are compared 
to the results obtained for symmetric matter ($\eta$=0)  
shown as dotted lines.  
\label{dwFT_Kstrbar_dens_rev1}
}
\end{center}
\end{figure}
\begin{figure}
\vspace{-1.4cm}
\hspace{-0.8cm}
\begin{center}
%\begin{tabular}{c c }
%\includegraphics[width=16cm,height=16cm]{dwFT_phi_dens_rev.eps}
\includegraphics[width=20cm,height=20cm]{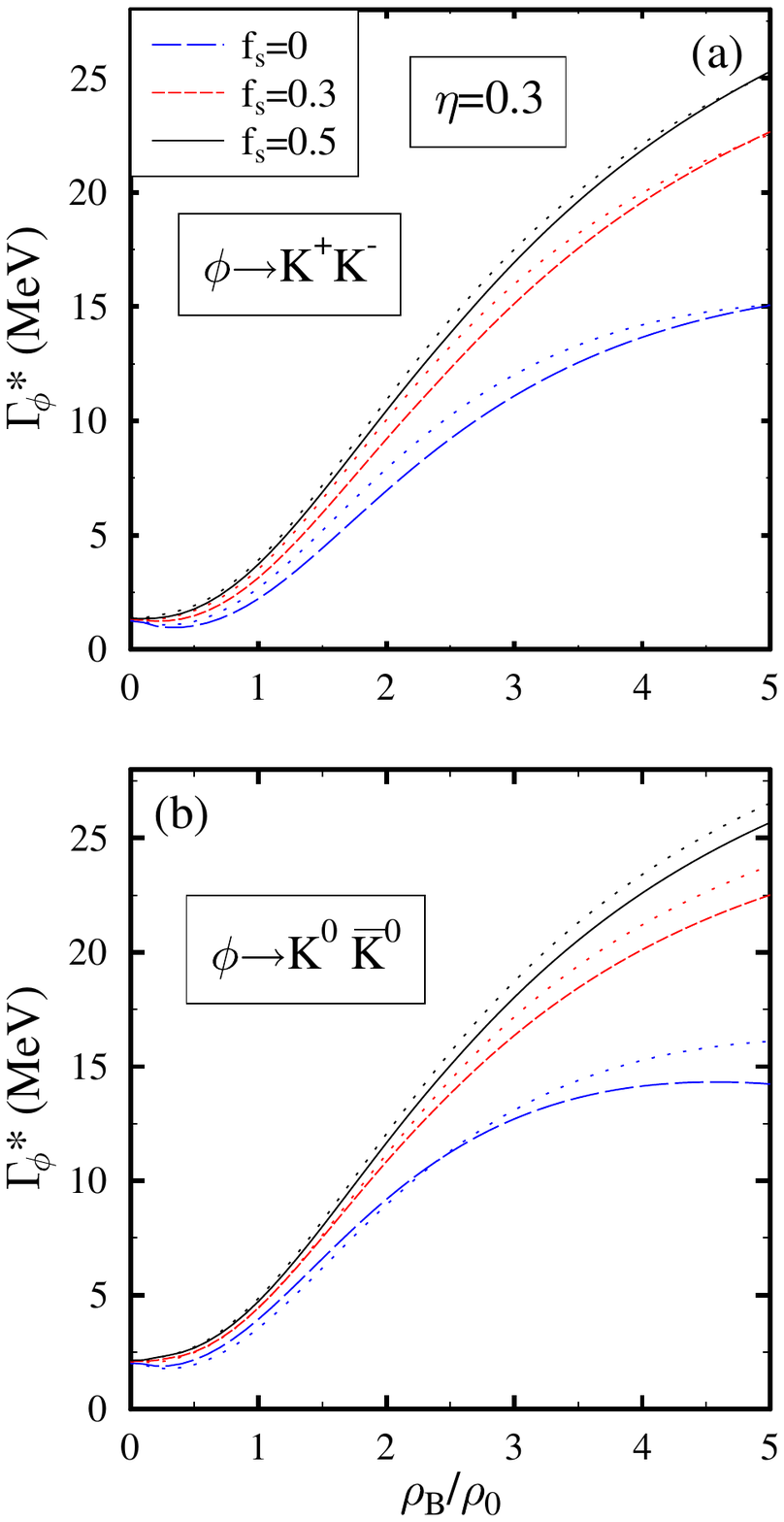}
\caption{
Decay widths of $\phi$ to charged and neutral $K \bar K$
in strange hadronic matter, for $f_s=0,0.3,0.5$ for
$\eta$=0.3. 
These are compared with the results for
isospin symmetric matter shown as dotted lines.
\label{dwFT_phi_dens_rev}
}
\end{center}
\end{figure}

\subsection{DECAY WIDTHS:}

The decay widths of $\phi \rightarrow K\bar K$ and 
$K^* (\bar {K^*}) \rightarrow K (\bar K) \pi$ 
are calculated using a field theoretic model of
composite hadrons \cite{spm781,spm782}. 
These are evaluated from the matrix element
of the light quark-antiquark pair creation term of the
free Dirac Hamiltonian between initial and final states 
\cite{amspmwg,amspm_upsilon,charmonium_PV_amspm,dmeson_PV_amspm,strange_PV_amspm}. 
For a generic decay process $A\rightarrow B +C$, the matrix element 
is multiplied with a parameter $\gamma_A$, which is a measure 
of the light quark-antiquark pair creation strength to form 
the final mesons $B$ and $C$ from the decay of the parent 
meson $A$. This parameter is fitted from the observed decay width
of $A\rightarrow B +C$ in vacuum.
This is similar to the calculation of the decay width
of $A\rightarrow B + C$ in the $^3P_0$ model \cite{3p0,3p0_1}, 
where the decay of the meson $A$ at rest 
to mesons $B$ and $C$ proceeds with
a light quark-antiquark pair creation in the $^3P_0$ state, and,
the matrix element is multiplied by a parameter, which is a 
measure of the quark-antiquark pair creation strength, 
fitted from the measured value of the decay width in vacuum.
The $^3P_0$ model has been extensively used to study the
decay processes of the mesons \cite{3p0,3p0_1,YOPR}
in vacuum. In the charm sector, the model is observed 
to explain the experimentally observed 
strong suppression of the decay modes 
of the $\psi(4040)$ to $D\bar D$ as well as 
$(D\bar {D^*}+\bar D D^*)$, as compared to the final state
$D^* \bar {D^*}$ in vaccum \cite{YOPR_psi4040}.
This suggests the importance of accounting for the
internal structure of the hadrons, i.e.,
the quark (and antiquark) constituents, 
to study their decay widths. 
The $^3P_0$ model has been used for the study of 
the in-medium decay widths
of the charmonium state to $D$ and $\bar D$ mesons 
from the mass modifications
of the $D$ and $\bar D$ mesons \cite{friman} as well as
from the in-medium masses of the charmonium as well as
open charm mesons calculated using a chiral effective model
\cite{amarvepja}. The process $\Psi \rightarrow D \bar D$
proceeds with $c(\bar c)$ of the charmonium state, $\Psi$
combining with the light antiquark (quark) of the quark-antiquark
created in the $^3P_0$ state, to form the $D$ and $\bar D$
mesons in the final state. Taking the internal structure of
the mesons in the initial and final states into account, 
along with the light constituent quark-antiquark pair creation
into account, both the models, e.g., the $^3P_0$ model as
well as the model of composite hadron as considered in the
present work, are observed to lead to vanishing of the
decay widths at specific densities \cite{friman,amarvepja,amspmwg}.

In the field theoretical model of composite hadrons considered
in the present work, we study the medium modifications of the
decay widths of $\phi \rightarrow K \bar K$ and
$K^* (\bar {K^*}) \rightarrow K (\bar K)\pi$ 
from the mass modifications of the open strange mesons.
These decay widths are calculated from the matrix element
of the free Dirac Hamiltonian between the initial and
final states, assuming the harmonic oscillator wave functions 
in the explicit constructions for these mesons
\cite{strange_PV_amspm}. The mesons $K$, $\bar K$, $K^*$, 
$\bar {K^*}$, $\pi$ and $\phi$ are assumed to be in the 1S state 
The model has been used to study the 
decay widths of charmonium state 
to $D\bar D$ as well as $D^* \rightarrow D\pi$ 
\cite{amspmwg} and of bottomonium states to $B\bar B$
in isospin asymmetric strange hadronic matter \cite{amspm_upsilon}.
The effect of strong magnetic fields has also been considered
on the charmonium decay widths to $D\bar D$ \cite{charmonium_PV_amspm},
$D^*\rightarrow D\pi$ \cite{dmeson_PV_amspm} as well as 
$K^*\rightarrow K\pi$ \cite{strange_PV_amspm},
where the pseudoscalar vector meson mixing effect is observed
to have significant contributions to the masses of these mesons, 
in addition to the Landau level contributions to the masses
of charged mesons.

\begin{figure}
\vspace{-1.4cm}
\hspace{-0.8cm}
%\begin{center}
%\begin{tabular}{c c }
%\includegraphics[width=16cm,height=16cm]{spectr_kstr.eps}
\includegraphics[width=18cm,height=18cm]{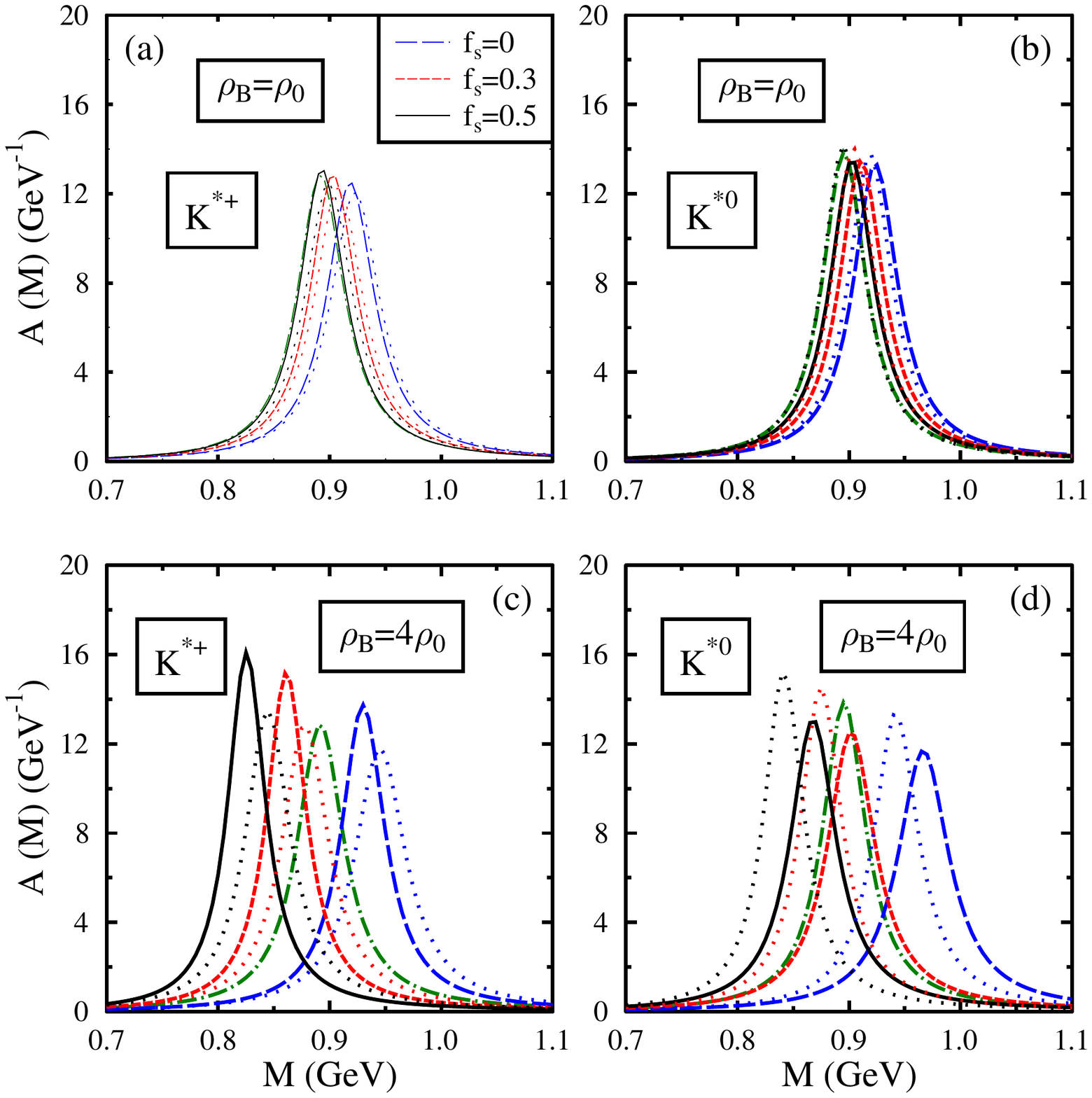}
\caption{
Spectral functions of the ${K^*}({K^*}^+,{K^*}^0)$ mesons
for $f_s=0,0.3,0.5$ for isospin asymmetric matter
with $\eta$=0.3.
These are compared with the results for
isospin symmetric matter shown as dotted lines.
The vacuum spectral function is shown as the dot-dashed line.
\label{spectr_kstr}
}
%\end{center}
\end{figure}
\begin{figure}
\vspace{-1.4cm}
\hspace{-0.8cm}
%\begin{center}
%\begin{tabular}{c c }
%\includegraphics[width=16cm,height=16cm]{spectr_kstrbar.eps}
\includegraphics[width=18cm,height=18cm]{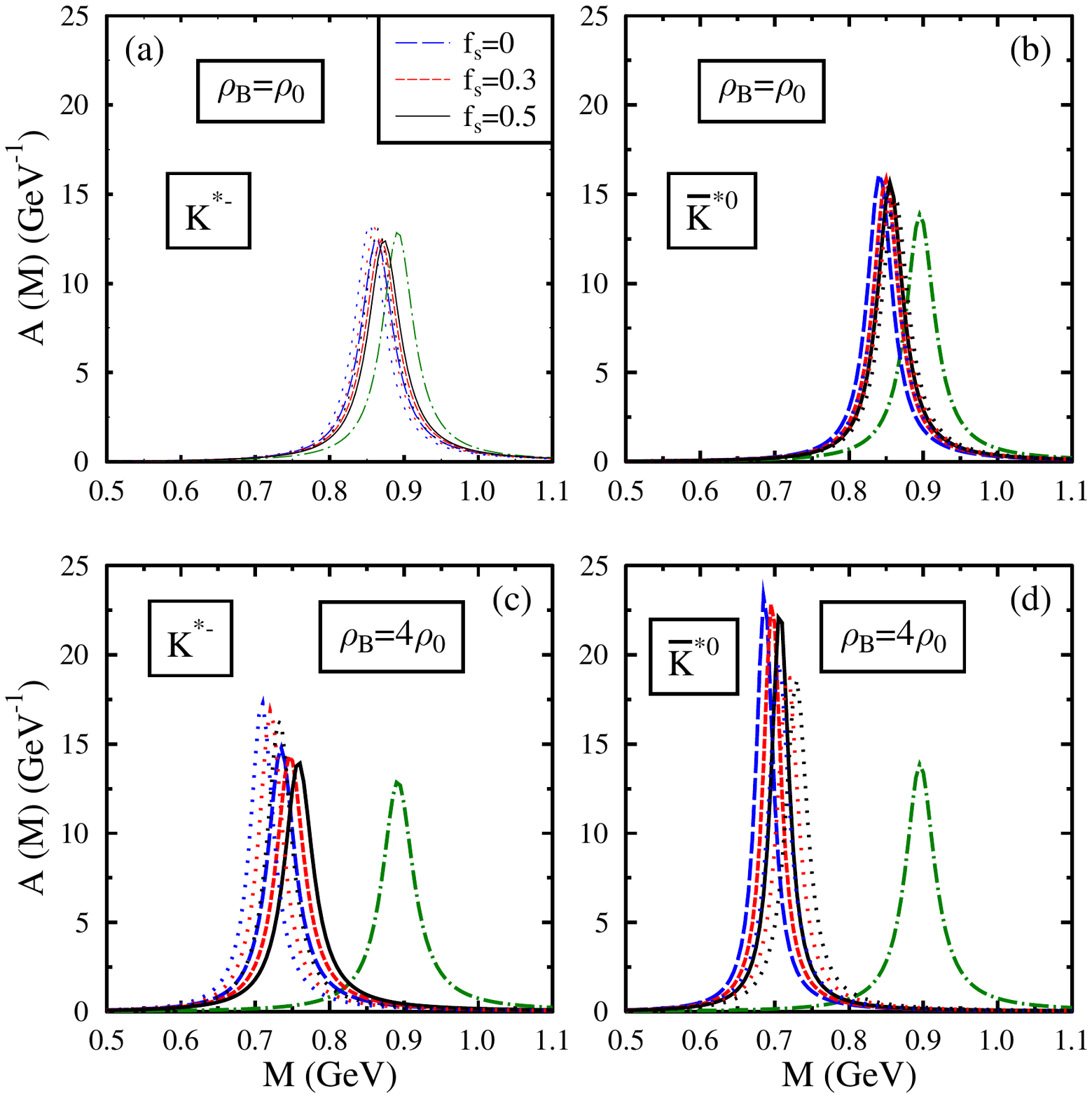}
\caption{
Spectral functions of the 
$\bar {K^*}({K^*}^-,\bar {{K^*}^0})$ mesons
for $f_s=0,0.3,0.5$ for isospin asymmetric matter
with $\eta$=0.3.
These are compared with the results for
isospin symmetric matter shown as dotted lines.
The vacuum spectral function is shown as the dot-dashed line.
\label{spectr_kstrbar}
}
%\end{center}
\end{figure}
The decay width of vector meson $A$ at rest
decaying to pseudoscalar mesons $B ({\bf p})$ and $C(-{\bf p})$ 
is given as
\begin{equation}
\Gamma\left(A\rightarrow B+C\right)
=\gamma_{A}^2 g^2\frac{8\pi^2p_B^0p_C^0}{3m_{A}}A_{A}
(|{\bf p}|)^2|{\bf p}|^3,
\label{gammaABC}
\end{equation}
where $p_B^0=(|{\bf p}|^2+m_B^2)^{1/2}$ and
$p_C^0=(|{\bf p}|^2+m_B^2)^{1/2}$ are the energies of
the outgoing $B$ and $C$ mesons respectively, in terms
of the magnitude of the 3-momentum of $B(C)$ meson,
$|{\bf p}|$, given as
\begin{equation}
|{\bf p}|=\Bigg (\frac{m_{A}^2}{4}-\frac{m_{B}^2+m_{C}^2}{2}
+\frac{\left(m_{B}^2-m_{C}^2\right)^2}{4m_{A}^2}
\Bigg )^{1/2}.
\label{momBC}
\end{equation}
For the decay $K^*(\bar {K^*})\rightarrow K (\bar K)\pi$,
$A_{K^*}(|{\bf p}|)$ is given as
\begin{eqnarray}
A_{K^*}(|{\bf p}|)=6c_{K^*}
\Big(\frac{\pi}{a_{K^*}}\Big)^{{3}/{2}}
\exp\Big[\Big (a_{K^*}b_{K^*}^2
-\frac{1}{2}\big(\lambda_2^2 R_K^2
+\frac{1}{4}R_\pi^2\big)\Big){|\bf p|}^2\Big] 
%\nonumber \\
\Big[{F_0}_{K^*}+\Big (\frac {3{F_1}_{K^*}}{2a_{K^*}}\Big )\Big],
\label{apkstr}
\end{eqnarray}
where,
\begin{eqnarray}
{F_0}_{K^*}&=&(b_{K^*}-1)\left(1-\frac{1}{8M_q^2}|{\bf p}|^2
(\lambda_2-\frac{1}{2})^2\right) \nonumber \\
&-&(b_{K^*}-\lambda_2)\left(\frac{1}{2}+\frac{1}{4M_q^2}
{|{\bf p}|}^2
\left(\frac{3}{4}b_{K^*}^2-\frac{5}{4}b_{K^*}
+\frac{7}{16}\right)\right)
\nonumber\\
&-&(b_{K^*}-\frac{1}{2})\Big[\frac{1}{2}+\frac{1}{4M_q^2}{|{\bf p}|}^2
\Big(\frac{3}{4}b_{K^*}^2-(1+\frac{1}{2}\lambda_2)b_{K^*}
+\lambda_2-\frac{1}{4}\lambda_2^2\Big)\Big]
\label{c0kstr}
\\
{F_1}_{K^*}&=&-\frac{1}{4M_q^2}\left[\frac{5}{2}b_{K^*}-\frac{9}{8}
-\frac{11}{12}\lambda_2\right].
\label{c1kstr}
\end{eqnarray}
The parameters $a_{K^*}$, $b_{K^*}$, $c_{K^*}$ are given as
\begin{eqnarray}
 a_{K^*}=\frac{\left(R_{K^*}^2+R_K^2+R_\pi^2\right)}{2},\;\;
b_{K^*}=\frac{1}{2a_{K^*}}\left(R_K^2\lambda_2
+\frac{1}{2}R_\pi^2\right),\;\;
 c_{K^*}=\frac{1}{12\sqrt 3}
\left(\frac{R_{K^*}^2 R_K^2 R_\pi^2}{\pi^3}\right)^{{3}/{4}}.
\label{abckstr}
\end{eqnarray}
In equations (\ref{c0kstr})--(\ref{c1kstr}), $M_q$ is the constituent
light quark ($q=u,d$) mass.
In equation (\ref{gammaABC}), for the decay of 
$K^* (\bar {K^*})\rightarrow K (\bar K)\pi$,
the value of $g^2$=1(2) for the neutral (charged) pion
in the final state.

For the decay process $\phi \rightarrow K \bar K$,
$A_{\phi}(|{\bf P}|)$ in equation (\ref{gammaABC}) is given as
\begin{eqnarray}
A_{\phi}(|{\bf P}|)= 6c_\phi
\Big(\frac{\pi}{a_\phi}\Big)^{{3}/{2}}
\exp[(a_\phi {b_\phi}^2
-R_K^2\lambda_2^2)|{\bf P}|^2]
\Big[F_{0 \phi}+F_{1 \phi}\frac{3}{2a_\phi} \Big],
\label{ap}
\end{eqnarray}
where $|{\bf P}|$ is the momentum of the outgoing $K(\bar K)$ meson
arising from the decay of $\phi$ meson at rest, which is
given by equation (\ref{momBC}), with particles (A,B,C) given by 
($\phi,K,\bar K$). 
In equation ({\ref{ap}),
\begin{eqnarray}
&&F^{\phi}_0 =  (\lambda_2-1)-\frac{1}{2M_q^2}
|{\bf P}|^2(b_{\phi}-\lambda_2)
\Big(\frac{3}{4}b_{\phi}^2-(1+\frac{1}{2}\lambda_2)b_{\phi}
+\lambda_2-\frac{1}{4}\lambda_2^2\Big),
\nonumber\\
&&F^{\phi}_1=\frac{1}{4 M_q^2}
\left[-\frac{5}{2}b_{\phi}+\frac{2}{3}
+\frac{11}{6}\lambda_2\right].
\label{c012phi}
\end{eqnarray}
and the parameters $a_\phi$, $b_\phi$ and $c_\phi$ given as
\begin{eqnarray}
a_\phi=\frac{1}{2}R_{\phi}^2+R_K^2, \;\;
b_\phi=R_K^2\lambda_2/a_\phi, \;\;
c_{\phi}=\frac{1}{6\sqrt{6}}\cdot
\left(\frac{R_\phi^2}{\pi}\right)^{{3}/{4}}
\cdot\left(\frac{R_K^2}{\pi}\right)
^{{3}/{2}},
\label{abcphi}
\end{eqnarray}

\subsection{Spectral functions and Production Cross-sections 
of vector mesons}

The relativistic Breit Wigner spectral function of
the vector meson $V$ is given as
\cite{Elena_16,Elena_17,Elena_13_1}

\begin{equation}
A_V (M)=C_1 \frac{2}{\pi} 
\frac {M^2 \Gamma_V^*}{(M^2-{m_V^*}^2)^2+(M\Gamma_V^*)^2},
\label{Spectral_V}
\end{equation}
where $M$ is the invariant mass and $C_1$ is a constant 
determined by normalization condition
\begin{equation}
\int _0 ^{\infty} A_V (M) d M=1.
\end{equation}
The in-medium relativistic Breit Wigner cross-section 
for the production of the vector meson, $V$
from the scattering of particles $a$ and $b$
%($ab\equiv K\pi$, $\bar K \pi$ and $K\bar K$ for the vector mesons
%$K^*$, $\bar {K^*}$ and $\phi$ respectively) 
is given as \cite{Elena_16,Elena_17,BW_CS_Haglin,BW_CS_Li}
\begin{equation}
\sigma (M)=\frac{6 \pi^2
\Gamma_V^* A_V (M)}{q(m_V^*,m_a^*,m_b^*)^2}.
\label{sigma1}
\end{equation}
In the above, $q(m_V^*,m_a^*,m_b^*)$
is the momentum of the scattering particle $a(b)$ in the 
center of mass frame of the vector meson, $V$, given as,
\begin{eqnarray}
q(m^*_V,m_a^*,m_b^*)
= \frac{1}{2M}
\Big (\big [{m_V^*}^2-(m_a^*+m_b^*)^2\big]
\times \big [{m_V^*}^2-(m_a^*-m_b^*)^2 \big] \Big )^{1/2},
\label{cm_mom}
\end{eqnarray}
with $m_V^*$, $m_a^*$ and  $m_b^*$ as the in-medium masses of
the vector meson, $V$ and the scattering particles
$a$ and $b$ respectively. 
The vector meson, $V$ may, however, be created from 
scattering of (as well as decay to) particles in different modes,
say particles $a_i$ and $b_i$ with masses $m^*_{a_i}$ 
and $m^*_{b_i}$ in the channel $i$. The in-medium decay width 
of the vector meson is then given as the sum of the decay 
widths in these channels, i.e., $\Gamma_V^*={\sum_i} {\Gamma_V^*}^i $.
The production cross-section of the vector meson, accounting for
all these channels is given as 
\begin{equation}
\sigma (M)=6 \pi^2 \Bigg (
\sum _i \frac{{\Gamma_V^*}^i}{q({m^{*}_V},m^*_{a_i},m^*_{b_i})^2}
\Bigg) A_V (M),
\label{sigmaV}
\end{equation}
where $q(m^*_V,m^*_{a_i},m^*_{b_i})$ is the
center of mass momentum of the particle $a_i$ as well as $b_i$ 
corresponding to the channel $i$ in the center of mass frame
of the vector meson, $V$. 
%As might be seen from equation 
%(\ref{cm_mom}), $q(m^*_V,m^*_{a_i},m^*_{b_i})$ depends on 
%the in-medium masses of the particles in channel 
%$i$ as well as on the in-medium mass of the vector meson, $V$. 

In the present work, we study the in-medium spectral functions 
as well as the production cross-sections of the vector mesons, 
$K^* ({K^*}^+, {K^*}^0)$, $\bar {K^*} ({K^*}^-, \bar {K^*}^0)$ 
and $\phi$ mesons in isospin asymmetric strange hadronic matter.
For the processes $K^* \rightarrow K\pi$, 
$\bar {K^*} \rightarrow {\bar  K}\pi$, 
and $\phi \rightarrow K\bar K$, the in-medium decay widths of
the vector mesons, $K^*$, $\bar {K^*}$ and $\phi$ 
are calculated 
from the medium modifications of the open strange mesons, 
using the field theoretic model of composite hadrons, as
described in the previous subsection.
The masses of the pseudoscalar strange mesons, $K$ and $\bar K$
mesons, as modified in the isospin asymmetric hyperonic matter 
are calculated using the chiral SU(3) model. 
The mass modifications of the $\phi$ meson and pions have not been
taken into account in the present work.
The shifts in the masses of the 
$K^*$ and $\bar {K^*}$ are assumed to be same as for the
mass shifts of $K$ and $\bar K$ mesons
\cite{qmc_phi,Krein_Prog_Part_Nucl_Phys}.
The in-medium spectral functions as well as the production 
cross-sections, given by equations 
(\ref{Spectral_V}) and (\ref{sigmaV}) respectively,
for the vector mesons, $K^*$, $\bar {K^*}$
and $\phi$ are investigated in the isospin asymmetric
strange hadronic matter.

\section{Results and Discussions}

The in-medium spectral functions for the hidden and open
strange vector mesons ($\phi$, $K^*$ and $\bar {K^*}$)
are investigated in the present work.
The decay widths of the vector mesons
for the decay processes $\phi \rightarrow K \bar K$ and
$K^* (\bar {K^*}) \rightarrow K (\bar K) \pi$ are calculated
using the field theoretic model of composite hadrons
\cite{strange_PV_amspm} as described in the previous section.
The medium modifications of these decay widths are 
computed from the mass modifications of the
open strange mesons in the hadronic medium.
The masses of the kaons and antikaons in the
isospin asymmetric strange hadronic medium 
have been calculated using a chiral SU(3) model 
in Ref. \cite{isoamss2}. 
The mass modifications of the $K$ and $\bar K$ mesons
arise due to the vectorial Weinberg--Tomozawa term
%%%%added below
at the leading order
%%%%added above
as well as the scalar exchange and the range terms 
at the sub-leading level 
%%%%added below
in the chiral perturbation theory.
%%%%added above
These are obtained by
solving the dispersion relation given by equation
(\ref{disp}) with the 
self-energies of the $K$ and $ \bar K$ mesons given by
equations (\ref{selfk}) and (\ref{selfkbar}) respectively. 
For the sake of completeness, the in-medium masses of the
kaons and antikaons, as calculated in Ref. \cite{isoamss2},
are shown in figure \ref{mkkbar_dens}, illustrating 
the effects due to the 
strangeness and isospin asymmetry of the hadronic medium.

For the isospin symmetric nuclear matter ($\eta$=0,$f_s$=0),
the Weinberg--Tomozawa term leads to a rise (drop) of the
mass of the $K (\bar K$) meson. The scalar exchange term
as well as the $d_1$ and $d_2$ range terms
are attractive, whereas, the first range term is repulsive
for both the kaons and antikaons.
In the presence of isospin asymmetry in the medium ($\eta\ne 0$), 
there are explicit contributions arising from the Weinberg-Tomozawa,
the terms involving $\delta$ in  the scalar exchange, the first 
range and the $d_2$ range terms. The $d_1$ range term also
implicitly depends on the isospin asymmetry, where the values
of the scalar densities of the baryons are obtained 
from the coupled equations of motion of the scalar fields,
$\sigma$, $\zeta$ and $\delta$. For nonzero strangeness
in the medium, there are contributions from the hyperons 
to the Weinberg-Tomozawa term and the $d_1$ and $d_2$ range terms.

As can be seen from figure \ref{mkkbar_dens}, the isospin asymmetry 
of the medium is observed to lead to a drop (increase) in the mass 
of the $K^+ (K^0)$ meson for the nuclear as well as in hyperonic
matter and these effects are observed to be larger for the
$K^0$ meson as compared to $K^+$ meson. In isospin symmetric nuclear
matter ($\eta$=0, $f_s$=0), the drop in the masses
of the antikaons in the medium as shown in figure \ref{mkkbar_dens} 
is due to the attractive contributions from the Weinberg-Tomozawa 
term, scalar exchange as well as the $d_1$ and $d_2$
range terms, which dominate over the repulsive contribution
from the first range term. There is observed to be a rise (drop)
of the mass of the $K^-$ ($\bar {K^0}$) meson mass due to
isospin asymmetry both in the nuclear as well as hyperonic matter.
The effects of isospin asymmetry 
lead to mass differences in the $K^+$ and $K^0$
as well as for $K^-$ and $\bar {K^0}$ mesons
(opposite in sign for the mass shifts
of $K^+$ and $K^0$ mesons in kaon doublet as well as 
of $K^-$ and $\bar {K^0}$ mesons in the antikaon doublet).
%%%%%%%modified below as per comment 3 of Referee 1
The mass shifts of around $+27$ and $-42$ MeV are obtained 
for $K$ and $\bar K$ mesons in isospin symmetric nuclear matter
at the nuclear matter saturation density
using the chiral SU(3) model \cite{isoamss2}.
%%%%%%%%%%%
%*******************************************************
These may be compared to the values of around $+29$ and $-50$ MeV 
\cite{tolos_2008} for these mesons calculated using 
a coupled channel approach
\cite{Oset_RamosNPA635_1998_99}.
%%%%%%%tolos_2008 Strange mesons in nuclear matter at finite temp
%%%%%%%%L. Tolos, D. Cabrera and A. Ramos, PRC78,045205 (2008)
%%%%%toloskbarn08.pdf in ~/amruta/strangedecaywidths
%%%%%Oset_Ramos_NPA635_1998_99
%%%Oset_RamosNPA635_1998_99.pdf in ~/amruta/strangedecaywidths
%************************************************************
%***********added below
In Ref. \cite{Oset_RamosNPA635_1998_99}, 
using the lowest order chiral Lagrangian as well as a cut-off
to regularize the loop integrals, the self-consistent solutions of
the coupled channel Lippmann-Schwinger equations for the s-wave 
meson-nucleon interaction in the $S=-1$ sector generates 
the $\Lambda(1405)$, and reproduces the $K^-p$ $\rightarrow$ 
$K^-p$, $\bar {K^0} n$, $\pi^0 \Lambda$, $\pi^0 \Sigma$, 
$\pi^-\Sigma^+$ and  $\pi^+\Sigma^-$ cross-sections.
%***********added above
%************************************************************
The $\bar KN$ scattering amplitude in nuclear matter at finite
tamperature is calculated within the coupled channel approach
in Ref. \cite{tolos_2008}. 
%($\bar K N$, $\pi \Sigma$, $\eta \Lambda$ and $K\Xi$  for I=0,)
%$\bar K N$, $\pi \Sigma$, $\pi \Lambda$ $\eta \Lambda$ and $K\Xi$ 
%for I=1) 
%by using the leading order meson-baryon interaction
%in the chiral Lagrangian. 
This is evaluated incorporating
the Pauli blocking, the dressing of the $\bar K$ (with s-
and p-wave contributions)
and $\pi$ meson as well as mean field binding of the nucleons 
and hyperons. The s- and p-wave contributions to the
$\bar K$ self energy are due to the Weinberg-Tomozawa term 
and the coupling of $\bar K$ to the hyperon-hole 
excitations respectively, whereas the self-energy of the pion arises 
due to coupling of pion to particle-hole (1p1h),
%***********************************************************
%%%modified/added below (with Ref.  A_Ramos_E_Oset_NPA_671_481_2000 
%%%added below
$\Delta$-hole ($\Delta$h) and two particle hole (2p2h) excitations. 
Due to Pauli blocking, the mass of $K^-$ is observed to increase
upto a density of around 0.1 $\rho$, which is observed
to decrease with further increase in the density.
When the $K^-$ is dressed with s- and p-waves, there is a
drop in the mass of the $K^-$, which is observed to
be a smaller drop with density, when the pion self energy
is also taken into account \cite{A_Ramos_E_Oset_NPA_671_481_2000}.
%***********************************************************
%%%modified/added above (with Ref.  A_Ramos_E_Oset_NPA_671_481_2000 
%%%added above
%%%
%(fig.6 of the paper
%A. Ramos, E. Oset, NPA671,481 (2000)
%K- in nuclear medium
%(9906016_Ramos_Oset_NPA_671_481_2000.pdf in ~/amruta/strangedecaywidths)
%%%

For the kaon-nucleon scattering, $KN$ is the only channel,
and a self-consistent calculation is observed to lead
to a positive mass shift of around 29 MeV for $K$ meson
in cold symmetric nuclear matter at nuclear matter saturation 
density, which is very close to the value of 25 MeV calculated 
using the $T\rho$ approximation.
%%%%%%%modified above as per comment 3 of Referee 1
%A similar value (of around 30 MeV) for the mass shift
%for the $K^*$ meson, as the $K$ meson, is obtained 
%from a self consistent solution of the G-matrix
%using the coupled channel approach 
%\cite{Oset_Ramos_2001,Oset_Ramos_2010}. 
%%%%%Oset_Ramos_NPA_679_2001_616_0005046.pdf
%%%%and
%%%%%%Oset_Ramos_arxiv_0905_0973_hep_ph.pdf
%%%in amruta/strangedecaywidths
%*************************************************
%*****added here***********************************
%1006.3645_Tolos_Cabrera_Molina_Ramos_Oset.pdf in 
%***************************************************************
For $\bar {K^*}$ meson, the mass shift 
as obtained from the real part of the self energy
($\sim Re \Pi (m_{\bar {K^*}},{\bf q}=0)/(2 m_{\bar {K^*}})$)
using a self consistent solution of the scattering amplitude
within the coupled channel approach 
is observed to be around $-50$ MeV
\cite{tolos_1006_3454,tolos_1006_3645},
which is similar to the mass shift of the $\bar K$ meson 
\cite{tolos_2008}.
For the $K^* N$ scattering, due to absence of any resonance
near to the threshold, similar to $KN$ scattering,
the low density $T\rho$ approximation is a good approximation
for calculation of the self energy of the $K^*$ meson.
The value of the mass shift for the $K^*$ meson 
of around +50 MeV at the nuclear matter saturation density 
obtained in the $T\rho$ approximation is modified to the 
value of around +40 MeV using the self-consistent calculation
of the scattering amplitude \cite{Elena_13_2}.
%%%%added below in response to comment 5 of Referee 1
%***************************************************
The coupling of $K^*N$ to $KN$ has been considered
in Ref. \cite{Khemchandani_et_al_PRD91_094008_2015},
which could arise from the decay process of 
$K^* \rightarrow K\pi$, followed by the pion decaying 
to a particle-hole pair, thus leading to 
the process $K^* N \rightarrow K N$. 
The $K^* N$-$KN$ coupling can modify 
the production of the $K$ and $K^*$ mesons
resulting from nuclear collision experiments.
However, no resonance formation arising 
due to the $KN$-${K^*}N$ interaction is observed
\cite{Khemchandani_et_al_PRD91_094008_2015}.
%***************************************************
%%%%added above in response to comment 5 of Referee 1
%%%%added below
It might be mentioned here that the sign convention for the
self energy in the dispersion relation of the $K$ and $\bar K$
meson used in the present work is opposite to the 
convention used in Refs. 
\cite{Oset_RamosNPA635_1998_99,A_Ramos_E_Oset_NPA_671_481_2000,tolos_2008,tolos_1006_3454,tolos_1006_3645,Oset_Ramos_2001,Oset_Ramos_2010}. 
%%%%added above
The mass modifications of the hadrons with identical 
light ($u$, $d$) quark (antiquark) constitutents, e.g., $K$ and $K^*$, 
are observed to be very similar within the QMC model 
\cite{Krein_Prog_Part_Nucl_Phys}. This is due to the
fact that these medium changes of the hadrons arise
dominantly from the scalar potentials of the light 
quark (antiquark) constituents.  
%%%%%%%%%%modifying here%%%%%%%%%%%%%%%%%%%%
%It might be noted here that the calculation of the mass modifications 
%of the $K^*$ and $\bar {K^*}$ mesons using the chiral effective model
%is beyond the scope of the present work. 
%The interactions of the open strange vector mesons, $K^*$ and
%$\bar {K^*}$ with the baryons, using the Lagrangian ${\cal L}_{BW}$,
%for $W=V$, turn out to be of the form 
%%$\bar \psi^i \gamma^\mu \psi^j V_\mu$, with $i\ne j$,
%which vanish due to the assumption 
%$\bar \psi_i \gamma^\mu \psi_j \rightarrow \delta_{ij} \delta^{\mu 0}
%\langle \bar \psi_i \gamma^ 0 \psi_i
%\rangle \equiv \delta_{ij} \delta^{\mu 0} \rho_i$,
%as given by equation (\ref{vec_scalar_densities}). 
%The only vector meson fields which have non-zero
%contributions from the baryon-meson interaction terms are the
%$\omega$, $\rho_0$ and $\phi$ mesons. 
In the present work, as has already been mentioned, 
we assume the mass shifts 
of $K^*(\bar {K^*})$ mesons in the isospin asymmetric strange 
hadronic matter to be the same as the mass shifts 
of the $K (\bar K)$ mesons as calculated using the chiral SU(3) model. 
This is motivated from the QMC model, where the mass shifts
of the hadrons with the same nonstrange light quark and antiquark 
constituents are almost identical, arising dominantly from
scalar potentials of these constituents 
\cite{Krein_Prog_Part_Nucl_Phys}. 
%%%%%%%%%%modifying here%%%%%%%%%%%%%%%%%%%%
The chiral SU(3) model used in the present work has been
recently used to study the mass and decay width of the
$\phi$ meson \cite{Arvind_phi}. 
The values of the in-medium  masses (in MeV) of $K^+$ ($K^-$) meson
are observed to be around 525 (456) and 547 (311)
for $\rho_B=\rho_0$ and $\rho_B=4\rho_0$ respectively
in symmetric nuclear matter ($\eta$=0, $f_s$=0) 
in Ref. \cite{isoamss2}, which are used in the present work. 
These values may be compared with the values of around 
524 (457) and 560 (312) MeV of Ref. \cite{Arvind_phi}
for the $K^+(K^-$) meson
for the same densities in symmetric nuclear matter. 
One thus observes the values of the masses of the $K^\pm$
in the present work to be very similar at $\rho_B=\rho_0$
with the values in Ref. \cite{Arvind_phi},
whereas, at the higher density of $\rho_B=4\rho_0$,
there is observd to be around 13 MeV difference 
in the $K^+$ mass (about $2\%$) from the value
in Ref. \cite{Arvind_phi}. 
This is due to the frozen glueball approximation 
(the expectation value of the dilaton field $\chi$ 
kept fixed at its vacuum value, $\chi_0$),
for the calculation of the scalar fields
($\sigma$, $\zeta$ and $\delta$), using which the kaon and
antikaon masses have been calculated in Ref. \cite{isoamss2}
and used in the present work.
On the other hand, these scalar fields are calculated accounting for
the medium dependence of the $\chi$ field \cite{amarvepja},
which are used for obtaining the in-medium  masses of
the $K$ and $\bar K$ mesons in Ref. \cite{Arvind_phi}, 
The mass of $K^-$ at the higher density of $\rho_B=4\rho_0$, 
however, is observed to be similar to the value in 
Ref. \cite{Arvind_phi}. This is due to the reason that
the $K^-$ mass has negative contributions
from the Weinberg-Tomozawa term, as well as, 
from the scalar fields ($\sigma$, $\zeta$ and $\delta$) 
(arising from the scalar exchange and (total) range terms), 
and the small modifications of these scalar fields 
including the medium modification of $\chi$ is observed 
to lead to only marginal modification 
to its mass in symmetric nuclear matter. 
In the presence of isospin asymmetry as well as
strangeness fraction in the medium,
the differences in the masses of kaons and antikaons 
in the present work from the values calculated in 
Ref. \cite{Arvind_phi}, are due to the fact that 
the masses used in the present work 
are obtained from the scalar fields ($\sigma$, $\zeta$
and $\delta$), calculated assuming the frozen glueball 
approximation, whereas the medium modifications of the dilaton
field, $\chi$ has been taken into account \cite{amarvepja} 
for the calculations of the scalar fields \cite{amarvepja} 
used for the study of the masses of the $K$ and $\bar K$ 
mesons in Ref. \cite{Arvind_phi}.
%%%%%%%%%%%%%added above

The modifications of the $\phi$ meson mass in hadronic medium
calculated from the self-energies of the 
$\phi$ meson due to kaon-antikaon loop \cite{Ko_phi}, 
from QCD sum rule calculations \cite{Hatsuda_Lee},
Quark meson coupling (QMC) model 
\cite{qmc_phi,phi_mass_decay_QMC}
as well as other studies \cite{Klingl_phi,Cabera_phi}
have been observed to be very small (at most
2-3\% at $\rho_B=\rho_0$). 
%%%%%%%%%%%%%added below
The mass modifications of the $\phi$ meson calculated
from the $K$-$\bar K$ loop,
using the in-medium masses of the kaons and antikaons
obtained within the chiral effective model \cite{Arvind_phi},
is observed to be small (of the order of 2-3 MeV) 
at $\rho_B=\rho_0$ in nuclear matter.
%%%%%%%%%%%%%added above
In the present work, the in-medium spectral function
as well as the production cross-section of the 
$\phi$ meson are studied without 
accounting for the mass modification 
of the $\phi$ meson. 

The decay widths $K^* (\bar {K^*})\rightarrow K (\bar K)\pi$
are computed using the field theoretic model of composite
hadrons described in the previous section. These decay
widths are obtained using equations 
(\ref{gammaABC})--(\ref{abckstr}),
with (A,B,C)=($K^*(\bar {K^*}),K(\bar K),\pi$). 
%%%%%%%%%%%%%%%%modified the sentences below
The parameter, $\lambda_2$ in the expression for 
the decay width of $K^* (\bar {K^*})\rightarrow K (\bar K)\pi$ 
is the fraction of the energy of the $K (\bar K)$ meson
carried by the constituent light (u,d) quark (antiquark). 
For the $K(\bar K$) at rest, the value of $\lambda_2$
is the fraction of the mass of the $K(\bar K)$ meson 
carried by the constituent light quark (antiquark).
In the present work, we take the constituent
masses of light (u,d) and strange quarks to be 
$M_{u,d}$=330 MeV \cite{amspmwg}, $M_s$=480 MeV and 
the vacuum masses (in MeV) for charged (neutral) kaons 
(and antikaons) are taken as 493.68 (497.61) \cite{pdg_2020}.
Assuming the binding energy of the meson 
shared by the quark (antquark) constituent to be inversely 
proportional to the quark (antiquark) mass
\cite{spm782}, the value of the fraction of energy
of the $K (\bar K$) carried by the light quark (antiquark),
$\lambda_2$ is obtained as 0.71 \cite{strange_PV_amspm}.
%%%%%%%%%%%%%%%%modified the sentences above
The harmonic oscillator strengths of the $K$ and
$\pi$ states are taken to be $R_K$=(238.30$\; {\rm MeV})^{-1}$,
$R_\pi$=(210.81 MeV)$^{-1}$ \cite{spm782,amspmwg} 
from the charge radii squared of pion and kaon to be
$(0.63\, {\rm fm})^2$ and $(0.56 \, {\rm fm})^2$
\cite{pdg_2020} respectively \cite{strange_PV_amspm}.
The value of $R_\phi$ is obtained from the observed decay width
of $\phi \rightarrow e^+e^-$ of 
%1.26377 
1.27 keV \cite{pdg_2020}
to be (290.71 MeV)$^{-1}$. 
The value of $R_{K^*}$ is calculated by assuming
that $R_{K^*}/R_K=R^{bag}_{K^*}/R^{bag}_K$, where 
$R^{bag}_{K^*}$ and $R^{bag}_K$ are the bag radii of the
$K^*$ and $K$ mesons of 0.74 fermis and 
%0.574 
0.57 fermis
calculated in the Quark meson coupling (QMC) model
\cite{Krein_Prog_Part_Nucl_Phys}. This yields the value
of $R_{K^*}$ to be 184.84 MeV.

The values of $\gamma_{K^*}$ for the decays 
${K^*}^{\pm}\rightarrow (K(\bar K)\pi)^{\pm}$ and
${K^*}^0 (\bar {K^*}^0)\rightarrow (K (\bar K) \pi)^0$
as fitted to their observed vacuum decay widths 
of 50.75 and 47.18 MeV \cite{pdg_2020}, 
are obtained as 2.44 and 
%2.335 
2.36 respectively. 
These yield the values of the
decay widths for the channels of the 
${K^*}^{\pm}$ meson to $K^{\pm}\pi^0$ and 
$ K^0 (\bar {K^0}) \pi^{\pm}$ to be
16.89 and 33.86 MeV and of decay widths 
of ${K^*}^0 (\bar {K^*}^0)$ to 
$K^0 (\bar {K^0})\pi^0$ and $K^{\pm} \pi^{\mp}$ 
to be 15.91 and 31.27 MeV respectively. 

In figure \ref{dwFT_Kstr_dens_rev1}, the in-medium
decay widths of the $K^* \rightarrow K\pi$
%$\Big ({K^*}^+\rightarrow (K\pi)^+ (K^+ \pi^0, K^0 \pi^+$)
%and ${K^*}^0\rightarrow (K\pi)^0 
%(K^0 \pi^0, K^+ \pi^-) \Big)$
are plotted as functions of the baryon density for $\eta$=0.3
and $f_s=0,0.3,0.5$ and the results are compared
with the isospin asymmetric case ($\eta$=0).
In panels (a) and (c), the decay widths of ${K^*}^+$ meson
are shown for the channels ${K^*}^+\rightarrow K^+ \pi^0$,
and ${K^*}^+\rightarrow K^0 \pi^+$ respectively,
and the total decay width of ${K^*}^+ \rightarrow (K\pi)^+$,
which is the sum of the two channels, is plotted in panel (e). 
Panels (b), (d) and (f) show the in-medium decay  
widths for the channels ${K^*}^0 \rightarrow K^0 \pi^0$,
${K^*}^0 \rightarrow K^+ \pi^-$, and,
of ${K^*}^0 \rightarrow (K\pi)^0$ (the sum of the two channels).
The decay width of ${K^*}^+$ to $K^0 \pi^+$
is observed to have much more pronounced effects from both 
the strangeness as well as isospin asymmetry
of the hadronic medium,
as compared to the decay width of ${K^*}^+$ to $K^+ \pi^0$. 
As might be observed from figure \ref{mkkbar_dens},
the isospin asymmetry leads to a drop (rise) of the 
mass of $K^+(K^0)$ meson (and of ${K^*}^+({K^*}^0)$ meson). 
This is observed
as a substantial decrease in the decay width of
${K^*}^+$ to $K^0 \pi^+$ as compared to
${K^*}^+$ to $K^+ \pi^0$  and an appreciable
rise in the decay width of 
${K^*}^0$ to $K^+ \pi^-$  as compared to
${K^*}^0$ to $K^0 \pi^0$ in the isospin asymmetric
matter as compared to the isospin symmetric ($\eta$=0)
case. The increase in the strangeness fraction 
in the hadronic medium is observed to lead to appreciable 
drop (increase) in the decay width of ${K^*}^+({K^*}^0)$ 
to $(K\pi)^+ ((K\pi)^0)$, as compared to the 
nuclear matter ($f_s$=0) at high densities
as can be observed from panels (e) and (f) in 
figure \ref{dwFT_Kstr_dens_rev1}. 
These medium modifications of the masses of the 
${K^*}^+$ and ${K^*}^0$ mesons should modify the
yield of these particles in the asymmetric heavy ion
collision experiments.

In figure \ref{dwFT_Kstrbar_dens_rev1},
the decay widths of $\bar {K^*}$ to $(\bar K \pi)$
are plotted for the value of the isospin asymmetry parameter,
$\eta$=0.3 and for typical values
of the strangeness fraction.
The results are compared to the isospin symmetric case.
The decay widths for ${K^*}^-$ to $K^-\pi^0$, ${\bar {K^0}}\pi^-$,
and the sum of these two channels are plotted in panels
(a), (c) and (e) respectively, whereas the decay widths
for ${\bar {K^*}}^0$ to ${\bar {K^0}}\pi^0$, ${K^-}\pi^+$, 
and the sum of these two channels are shown in panels
(b), (d) and (f) respectively.
The effects from the strangeness fraction
on the decay widths of ${\bar {K^*}} \rightarrow {\bar K} \pi$ 
are observed to be rather moderate as compared to the effects 
on the $K^* \rightarrow K\pi$ decay widths
shown in figure \ref{dwFT_Kstr_dens_rev1}.
There is observed to be a large increase in the decay width
of the ${K^*}^-\rightarrow \bar {K^0} \pi^-$ as compared to
that of ${K^*}^-\rightarrow {K^-} \pi^0$, and a drop
in the decay width of the $\bar {K^*}^0\rightarrow \bar {K^-} \pi^+$ 
as compared to that of $\bar {K^*}^0\rightarrow \bar {K^0} \pi^0$
for the isospin asymmetric matter as compared to $\eta$=0 case.
This is because of an increase (drop) of the mass of
$\bar {K^0}$ ($K^-$) (as well as for $\bar {{K^*}^0}$ (${K^*}^-$)
with isospin asymmetry in the medium.
%%%%%%%%%%%%%%%%%modifying now
In symmetric nuclear matter, the total decay width of 
${K^*}^- \rightarrow (\bar K \pi)^-$ as well as of 
${\bar {K^*}}^0 \rightarrow (\bar K \pi)^0$
are observed to drop with increase in density.
The values of these decay widths for $\rho_B=\rho_0 (4 \rho_0)$
are obtained to be around 49.27 (37.67) and 43.28 (27.47)
in symmetric nuclear matter. These results may be compared
with the results of an appreciable increase
in the decay width of $\bar {K^*}$
(around five times the vacuum value
at the nuclear matter density),
obtained within a coupled channel approach 
\cite{tolos_1006_3454,tolos_1006_3645}.
The increase in the decay width 
within the coupled channel approach 
is due to the absorption channels 
$\bar {K^*}N \rightarrow \rho Y, \omega Y,
\phi Y, \; Y=\Lambda,\Sigma$
\cite{tolos_1006_3454,tolos_1006_3645}, 
in addition to the elastic channel
$\bar {K^*} N \rightarrow \bar {K^*} N$
as well as accounting for the decay channel
of $\bar {K^*} \rightarrow \bar K \pi$.
%%%%%%%%%%%%%%%%%modifying now
The pion created from the decay $\bar {K^*}\rightarrow K\pi$,
for example, can become a nucleon-hole pair
in the medium, leading to the additional process
of $\bar {K^*} N \rightarrow \bar K N $.
Though the mass drop is observed to be moderate 
($\sim$ 50 MeV) for ${\bar {K^*}}$ meson
in the coupled channel approach,
there is appreciable increase in its decay width 
due to coupling of ${\bar {K^*}}N$ to the 
hyperon-vector meson excitations,
which can further increase due to processes
$\bar {K^*} N \rightarrow \bar K N $.
%%%%%%added  below in response to comment 5 of Ref. 1
The in-medium decay width of $K^* (\bar {K^*})\rightarrow K(\bar K) \pi$,
which is the dominant decay channel of $K^* (\bar {K^*})$ meson,
is investigated in the present work from the mass modifications 
of the open strange vector and pseudoscalar mesons, but
the mass modifications of the pion, which is observed
to be small \cite{tolos_2008}, has not been taken into account
in the present work.
The coupling of ${\bar {K^*}} N$ to the various coupled channels 
as mentioned above, which lead to appreciable increase 
in the decay width of the $\bar {K^*}$
meson \cite{tolos_1006_3454,tolos_1006_3645}, 
are, however, beyond the scope of the mean field approach 
of the present work.
%%%%%%added  above in response to comment 5 of Ref. 1

We next consider the medium modification of the
decay width of $\phi \rightarrow K\bar K$. 
The values of $\gamma_\phi$ fitted from the
observed decay widths of $\phi$ to the final states $K^+K^-$
and $K^0 \bar {K^0}$ in vacuum of 2.09 and 1.44 MeV 
are obtained as 2.38 and 2.34 respectively.
In figure \ref{dwFT_phi_dens_rev}, the decay widths
of $\phi$ meson to $K^+K^-$ and $K^0 {\bar {K^0}}$
are plotted in panels (a) and (b) for different values of
the strangeness fraction, both for isospin symmetric
($\eta$=0) as well as isospin asymmetric matter 
(with $\eta$=0.3). 
At the nuclear matter density, the decay width 
%is observed to be 2.687, 3.515 and 3.9 MeV for the channel 
is observed to be 2.69, 3.52 and 3.90 MeV for the channel 
(a) $\phi\rightarrow K^+K^-$, and, 
%3.548, 4.43 and 4.85 for the channel 
3.55, 4.43 and 4.85 for the channel 
(b) $\phi\rightarrow K^0 {\bar {K^0}}$ for the values
of the strangeness fraction as $f_s$=0, 0.3 and
0.5 respectively. The value of the total width of
%6.235 MeV at $\rho_B=\rho_0$ 
6.24 MeV at $\rho_B=\rho_0$ 
in symmetric nuclear matter in the present work may
be compared with the value of around 25 MeV in Ref. \cite{Ko_phi},
a value of around 33-38 MeV (for the values of the momentum
cut-off parameter chosen) 
within QMC model \cite{phi_mass_decay_QMC},
%around 45 MeV in Ref. \cite{Klingl_phi}
and, the value of around 22 MeV 
using a coupled channel approach \cite{Oset_Ramos_2001}, 
where the additional channels $\phi N \rightarrow K Y$,
$Y=\Lambda,\Sigma$, as well as, $\phi N \rightarrow \Sigma^* K$,
with $\Sigma^* \equiv \Sigma (1385)$ resonance,
are incorporated through the p-wave self energy of $\bar K$ 
due to the coupling to the hyperon-hole ($Y N^{-1}$)
states, as well as, to the $\Sigma^* N^{-1}$ state.
%%%%%%%%%%%%%%%%%%%%%%%%%%%%%%%%%%%%%%%%%%%%%%%%%
%%%%%%%%added below
The in-medium decay widths of the $\phi$ mesons
have also been studied experimentally 
from the study of the $\phi$ transperancy ratio
of the photoprduction data at Spring8-LEPS \cite{Spring8-LEPS}
and CLAS \cite{CLAS} collaborations as well as the
proton induced reactions in nuclei at COSY-ANKE 
\cite{M_Hartmann_PRC85_035206_2012}.
In Ref.  \cite{M_Hartmann_PRC85_035206_2012}.
the $\phi$ mesons produced by collisions
of protons with beam energy 2.83 GeV on C,Cu,Ag,Au targets
detected via $\phi$ decaying to $K^+ K^-$ at ANKE detector at COSY
were studied. The decay width of $\phi\rightarrow K^+K^-$
extracted from the nuclear tranperancy ratio showed 
an appreciably large value (of the order of 33-50 MeV)
of the in-medium decay wdith of $\phi\rightarrow K^+K^-$. 
%%%%%%%%added above
%%%%%%%%%%%%%%%%%%%%%%%%%%%%%%%%%%%%%%%%%%%%%%%%%
%%%%%%%%added below
In the present work of the composite model of hadrons,
the values obtained for the decay width (in MeV)
of the $\phi\rightarrow K\bar K$ 
(sum of the channels with the neutral and charged 
$K \bar {K}$) final states) are around 6.24 and 25.11 
in symmetric nuclear matter for $\rho_B=\rho_0$ 
and $\rho_B=3\rho_0$ respectively, which are very similar 
to the the values calculated in Ref. \cite{Arvind_phi} 
from the $K\bar K$ loop using the phenomenological interaction 
($\sim \phi^\mu (\bar K (\partial_\mu K)
-(\partial_\mu \bar K) K$) \cite{Ko_phi,phi_mass_decay_QMC}. 
The value of the decay width of $\phi\rightarrow K\bar K$ 
of around 6.24 MeV at the nuclear matter saturation density
in the present work as well as in Ref. \cite{Arvind_phi}
is thus observed to be much smaller than the value 
of around 22 MeV in the coupled channel approach
\cite{Oset_Ramos_2001}. As has already been mentioned,
the large value of the $\phi$ decay width arises
due to the coupling $\phi N$ to the 
additional channels $\phi N \rightarrow K Y$, 
$Y=\Lambda,\Sigma$, as well as, 
$\phi N \rightarrow \Sigma^* K$, which is beyond the
scope of the present work.
%%%%%%%%%%%%%%%%%%%%%%%%%%%%%%%%%%%%%%%%%%%%%%%%%
%%%%%%%%added above
%%%%%%%%%%%%%%%%%%%%%%%%%%%%%%%%%%%%%%%%%%%%%%%%%%

In the present work, as can be seen from figure 
\ref{dwFT_phi_dens_rev}, the increase in the strangeness 
fraction of the hadronic medium
leads to larger values of the decay widths 
of $\phi\rightarrow K\bar K$ (for both
the channels (a) $\phi \rightarrow K^+K^-$ as well as
(b) $\phi\rightarrow K^0 {\bar {K^0}}$). The effects of
isospin asymmetry on the $\phi$ meson decay widths
are observed to be rather moderate. However, the
effect of the strangeness which is seen to lead 
to increase in the $\phi$ meson decay width
and the rise in the decay width is observed 
to be quite pronounced at high densities. 
This should have observable consequences in the heavy ion collisions
at the Compressed baryonic matter (CBM)
experiments at future facility at GSI,
e.g., the production of $K$ and $\bar K$ mesons
to be more abundant as well as suppression of the $\phi$ meson.
A drop in the production Breit-Wigner cross-sections
of the $\phi$ meson (through the scattering of 
$K$ and $\bar K$ mesons) 
with the increase in the strangeness fraction of the medium, 
is observed, as might be seen from figure \ref{spectr_phi}.

\begin{figure}
\vspace{-1.4cm}
\hspace{-0.8cm}
%\begin{center}
%\begin{tabular}{c c }
%\includegraphics[width=16cm,height=16cm]{sigma_kstr.eps}
\includegraphics[width=18cm,height=18cm]{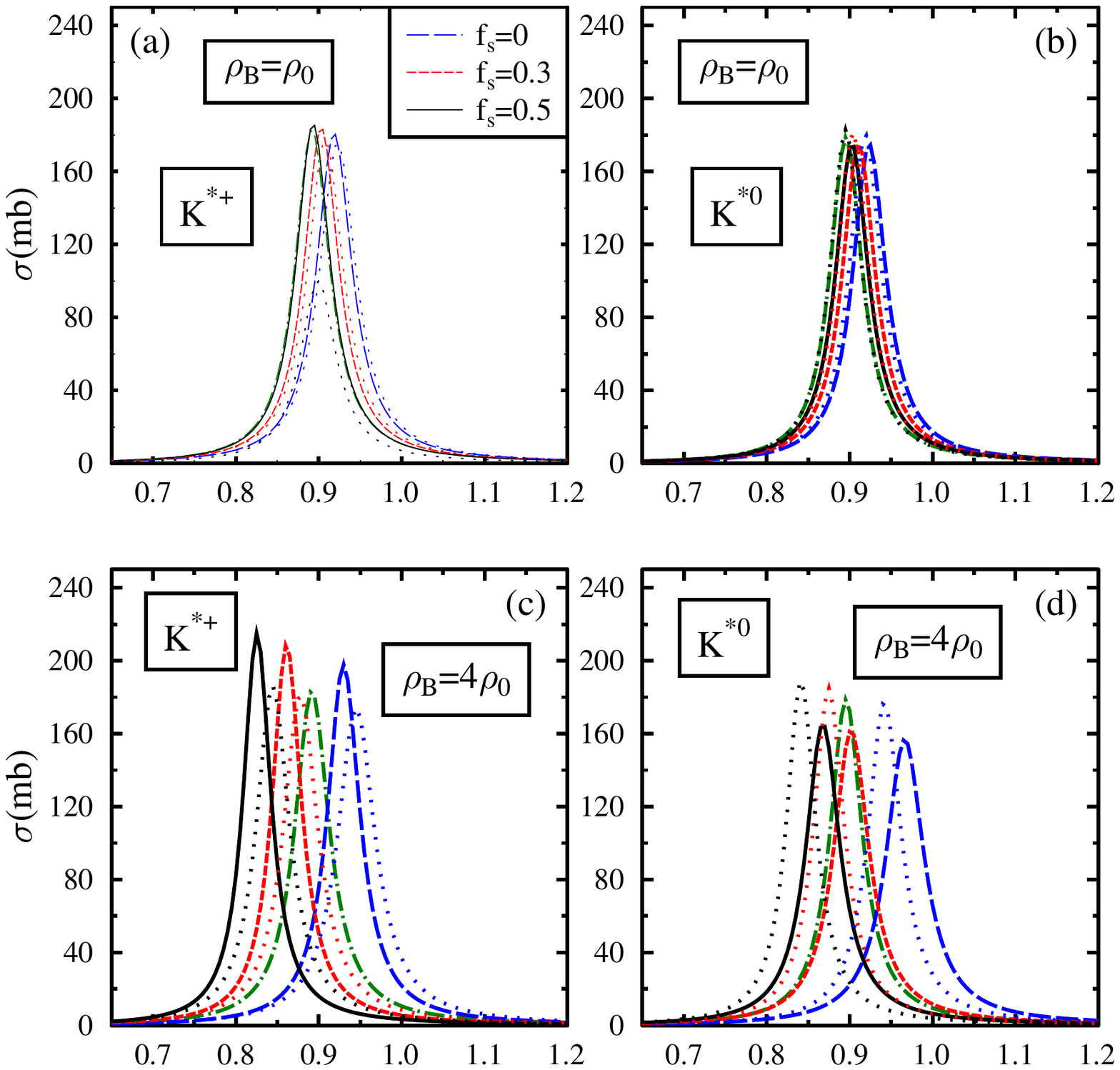}
\caption{
Production Cross-section of the 
${K^*}({K^*}^+,{K^*}^0)$ mesons
for $f_s=0,0.3,0.5$ for isospin asymmetric matter
with $\eta$=0.3. These are compared with results for 
symmetric matter ($\eta$=0) shown as dotted lines.  
The production cross-section for the vacuum case is shown 
as the dot-dashed line.
\label{sigma_kstr}
}
%\end{center}
\end{figure}
\begin{figure}
\vspace{-1.4cm}
\hspace{-0.8cm}
%\begin{center}
%\begin{tabular}{c c }
%\includegraphics[width=17cm,height=17cm]{sigma_kstrbar.eps}
\includegraphics[width=18cm,height=18cm]{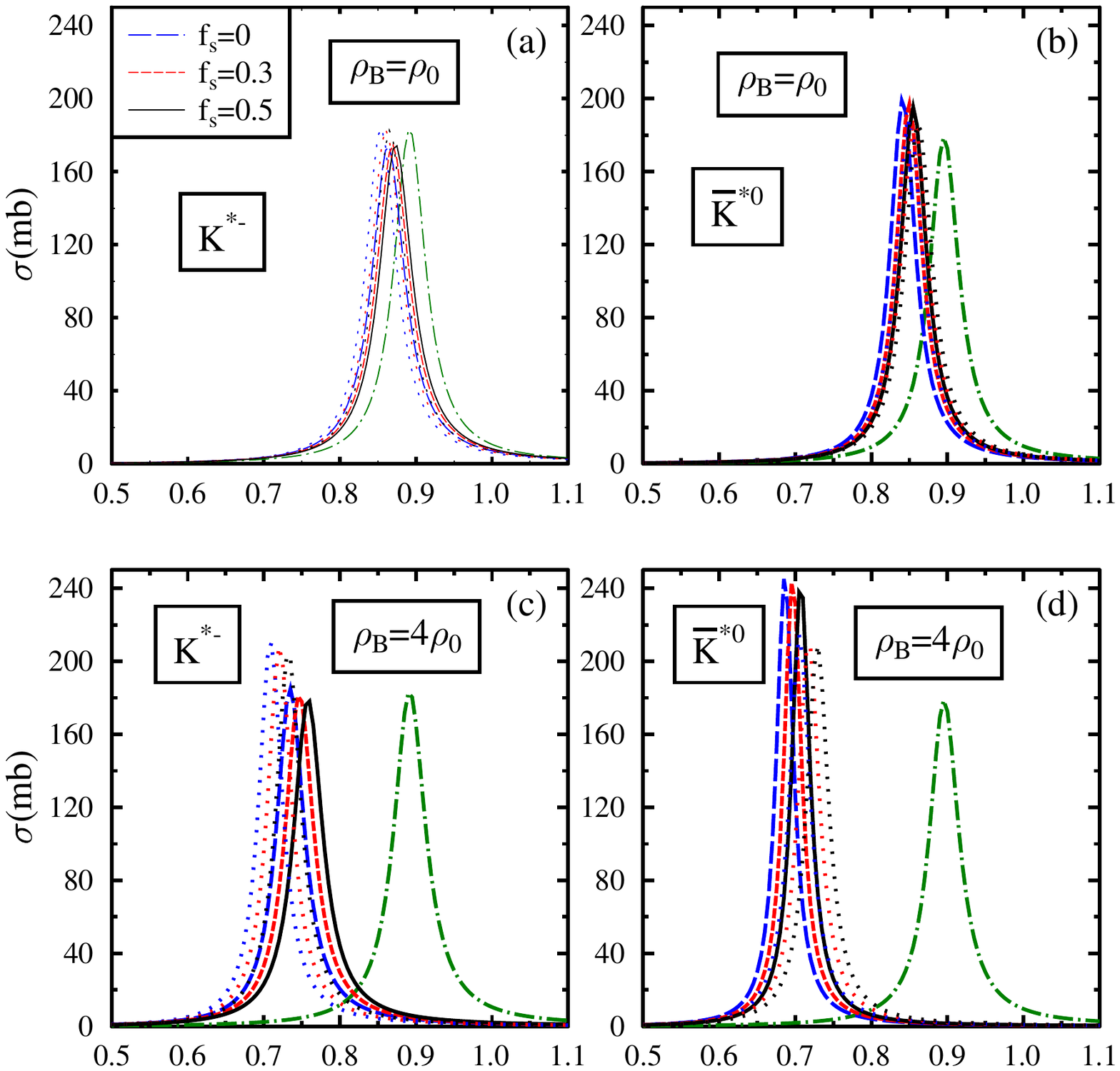}
\caption{
Production Cross-section of the
$\bar {K^*}({K^*}^-,\bar {{K^*}^0})$ mesons
for $f_s=0,0.3,0.5$ for isospin asymmetric matter
with $\eta$=0.3 and compared with results for
symmetric matter shown as dotted lines.
The production cross-section for the vacuum case is shown 
as the dot-dashed line.
\label{sigma_kstrbar}
}
%\end{center}
\end{figure}
\begin{figure}
\vspace{-1.4cm}
\hspace{-0.8cm}
%\begin{center}
%\begin{tabular}{c c }
%\includegraphics[width=16cm,height=16cm]{spectr_phi.eps}
\includegraphics[width=18cm,height=18cm]{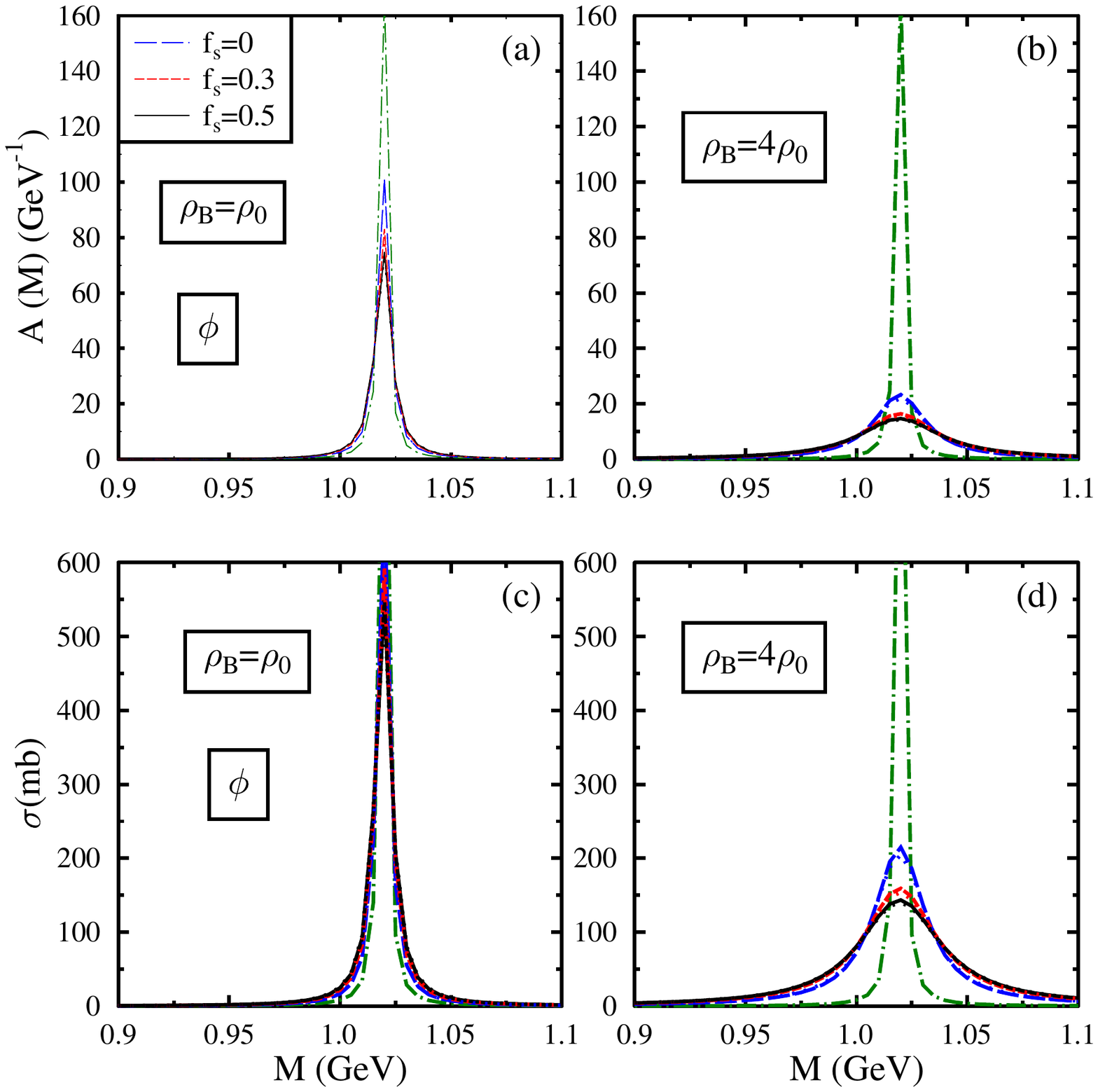}
\caption{
Spectral functions and Production Cross-section 
of the $\phi$ mesons
for $f_s=0,0.3,0.5$ for isospin asymmetric matter
with $\eta$=0.3 and compared with results for symmetric matter 
($\eta$=0) shown as dotted lines.  
The spectral function as well as the production cross-section
of $\phi$ meson for the vacuum case are shown 
as the dot-dashed lines.
\label{spectr_phi}
}
%\end{center}
\end{figure}

In figures \ref{spectr_kstr} and \ref{spectr_kstrbar},
we show the effects of the strangeness and
isospin asymmetry on the spectral functions of
the $K^*({K^*}^+,{K^*}^0$) and
$\bar {K^*}({K^*}^-,\bar {K^*}^0)$ 
mesons for densities $\rho_B=\rho_0$ and $\rho_B=4\rho_0$. 
In these figures, the vacuum spectral functions are shown 
as the dot-dashed curves.
For nuclear matter ($f_s$=0), there is observed to
be a shift in the peak to a higher invariant mass
for $K^*$ mesons from the vacuum value,
whereas there is a downward shift 
as the strangeness fraction is increased. The isospin 
asymmetry leads to a higher (lower) value for the position 
of the peak for the ${K^*}^0$ (${K^*}^+$) spectral function.
The observed behaviour is due to the assumption of
the mass shifts of the $K^*$ mesons to be same as 
the mass shifts of the $K$ meson as calculated in the
chiral SU(3) model. The spectral functions of the
$\bar {K^*} ({K^*}^-, \bar {K^*}^0)$ plotted
in figure \ref{spectr_kstrbar} show a much larger
downward shift of the peak position for the higher value of
density, $\rho_B=4 \rho_0$ as compared to $\rho_B=\rho_0$. 
The dependence on the isospin asymmetry as well
as strangeness fraction are observed to be much more dominant
for the ${K^*}^+$ and ${K^*}^0$ mesons as compared to the
${K^*}^-$ and $\bar {K^*}^0$ mesons.
The spectral functions of the $K^*$ and $\bar {K^*}$ 
have been studied using the coupled channel approach
in Refs. \cite{Elena_17,tolos_1006_3645}.
For $\bar {K^*}$ meson, there is observed to be appreciable
broadening of the quasi-particle peak due to the large
increase in its decay width arising from the absorption
channels ${\bar {K^*}} N \rightarrow VY, V=\rho,\omega,\phi,
Y=\Lambda,\Sigma$. There are two resonances  generated
dynamically, which can be identified with $\Lambda (1800)$
and $\Sigma (1750)$. The quasi-particle peak is
moved to much lower energy due to the interference of the
$\bar {K^*}$ with the resonance-hole excitations.
In the absence of the absorption channels, the 
quasi-particle peak remains narrow for $K^*$ meson,
which has contribution predominantly from the decay channel
$K^* \rightarrow K\pi$.

Figures \ref{sigma_kstr} and \ref{sigma_kstrbar},
the production cross-sections for the $K^*$ and
$\bar {K^*}$, as created from $K \pi$ and $\bar K \pi$ scatterings,
are plotted. These are shown for densities $\rho_0$ and $4\rho_0$.
In the asymmetric hadronic matter with $\eta$=0.3,
there is observed to be a modification in the maximum 
value of the production cross-section
of ${K^*}^+$ (${K^*}^0)$ meson from the vacuum value of
183 mb to around 180 (174), 183 (175) and  185 (174) 
for $\rho_B=\rho_0$ and 198 (156), 208 (161)
and 215 (164) for $\rho_B=4\rho_0$, 
for $f_s$=0, 0.3 and 0.5 respectively.
There are observed to be marginal changes from the 
vacuum case for the maximum cross-section 
for ${K^*}^+$ and ${K^*}^0$ mesons 
for symmetric hadronic matter ($\eta$=0)
shown as the dotted lines for both $\rho_0$ and $4\rho_0$.
For the ${K^*}^-$ meson, the maximum value for the
production cross-section (in milibarns) is observed 
to have small modifications
from the vacuum value of 182 to around
174 and 185 in asymmetric nuclear matter (with $\eta$=0.3)
for densities $\rho_0$ as well as
$4\rho_0$. However, the peak position for its production
is at the lower invariant mass (in GeV) of around 0.85 
and 0.74 for these densities. In the absence of isospin asymmetry
in the hadronic medium, the maximum values of the 
cross-sections are observed to be larger (smaller)
than the isospin asymmetric case for ${K^*}^-$ ($\bar {K^*}^0$)
mesons respectively. The maximum cross-section for $\bar {K^*}^0$
is observed to be around 199 mb at nuclear matter saturation
density and around 245 mb at $\rho_B=4\rho_0$ in asymmetric
nuclear matter. The modifications of $\sigma_{max}$ is observed to 
be marginal with finite strangeness fraction in the medium.
The production cross-sections of the ${K^*}^+$ as compared to
${K^*}^0$ meson from $K\pi$ scattering as well as
of ${\bar {K^*}^0}$ as compared to ${K^*}^-$ 
from $\bar K\pi$ scattering
in the isospin asymmetric hadronic medium at high densities, 
should show in the ${K^*}^+/{K^*}^0$ and 
${\bar {K^*}^0}/{K^*}^-$ ratios in the asymmetric 
heavy ion collision in the compressed baryonic experiment
at the future facility at GSI.

Figure \ref{spectr_phi} shows the spectral function 
as well as the production cross-section of $\phi$ meson 
due to $K\bar K$ scattering. One observes a broadening
of the peak for the higher density of $4\rho_0$ as
compared to the nuclear matter density.
The maximum value of the production cross-section 
(in milibarns) is observed to have a substantial reduction
to around 205 (215), 154 (159) and 140 (144) 
for $f_s$=0, 0.3 and 0.5 respectively at the higher baryon density
of $4\rho_0$ for hadronic matter with $\eta$= 0 (0.3). 
This should be observed as suppression of $\phi$ meson 
in the hadronic medium at high densities.

\section{Summary}

We have investigated the in-medium spectral functions
as well as the production cross-sections of strange
vector mesons ($K^*$, $\bar {K^*}$, $\phi$) 
in isospin asymmetric nuclear (hyperonic) matter.
The in-medium decay widths of these mesons, for 
the decay processes $ K^* \rightarrow K \pi $,
$\bar {K^*} \rightarrow {\bar  K} \pi$,
and $\phi \rightarrow K\bar K$ are 
computed from the medium modifications of 
the masses of the open strange mesons,
using a field theoretic model of composite
hadrons with quark (antiquark) constituents.
These are calculated from the matrix element of the
quark-antiquark pair creation term of the
free Dirac Hamiltonian between the inital and final states. 
The masses of the kaons and antikaons are calculated
using a chiral SU(3) model and the mass shift
of the vector $K^* (\bar {K^*})$ meson 
is assumed to be the same as the mass shift of the
$K (\bar K)$ meson in the hadronic medium. 
There is observed to be dominant contributions
to the kaon and antikaon masses
from the isospin asymmetry as well as strangeness
at high densities, which is observed to lead to 
large mass difference of the $K^+$ and $K^0$ mesons.
This should show in the experimental observable, e.g, 
the $K^+/K^0$ ratio in asymmetric heavy ion collision
expertiments.
There are observed to be significant modifications
of the decay widths of the $K^* ({K^*}^+,{K^*}^0)$,
and ${\bar {K^*}} ({K^*}^-,{\bar {K^*}}^0)$, from the
decay processes $K^*\rightarrow K\pi$ and
${\bar K^*}\rightarrow {\bar K}\pi$
in the isospin asymmetric strange hadronic matter
and the decay width of $\phi \rightarrow K \bar K$
is observed to be appreciably larger with increase
in strangeness in the medium. 

The effects of the strangeness and isospin asymmetry
on the spectral functions as well as the production
cross-sections of the $K^*$, $\bar {K^*}$ and $\phi$
mesons arising from the $K\pi$, $\bar K\pi$ and $K\bar K$
scatterings respectively, should show in the yields 
of these particles.
The substantial increase in the production cross-section
of ${K^*}^+$ as compared to ${K^*}^0$ meson,
of ${\bar {K^*}^0}$ as compared to ${K^*}^-$,
as well as of $\phi$ in the isospin asymmetric 
hadronic medium at high densities, 
should show in the yields of these
particles in asymmetric heavy ion collisions
at the compressed baryonic matter (CBM) experiments 
at the future facility at GSI.

One of the authors (AM) acknowledeges financial support 
from Department of Science and Technology (DST), 
Government of India (project no. CRG/2018/002226).

\end{document}